\documentclass[
aps,
twocolumn,
longbibliography,
nofootinbib
]{revtex4-2}
\usepackage[T1]{fontenc}
\usepackage{amsmath}
\usepackage{txfonts}
\usepackage{bm}
\usepackage{color}
\usepackage[normalem]{ulem}
\usepackage{tikz}
\usetikzlibrary{arrows.meta}
\usepackage[colorlinks=true,linkcolor=blue,citecolor=blue,urlcolor=blue,pdfpagemode=UseNone,pdfstartview=FitH]{hyperref}

\newcommand{\EJ}{\mathbb{E}_J}
\newlength{\mainparindent}
\newenvironment{romanlist}{\setlength{\mainparindent}{\parindent}\begin{enumerate}\setlength{\parindent}{\mainparindent}}{\end{enumerate}}

\makeatletter
\newcommand{\protectedsectionskip}{%
  \par\addpenalty\@secpenalty
  \dimen@\prevdepth\hrule\@height\z@\nobreak
  \vskip 0.8cm \@plus1ex \@minus.2ex\nobreak
  \prevdepth\dimen@\@nobreaktrue}
\makeatother
\begin{document}

\title{Physical interpretation of spin-glass models with correlated disorder}
\author{Hidetoshi Nishimori}
\affiliation{Institute of Integrated Research, Institute of Science Tokyo, Nagatsuta-cho, Midori-ku, Yokohama 226-8501, Japan}
\affiliation{Graduate School of Information Sciences, Tohoku University, Sendai 980-8579, Japan}

\begin{abstract}
Ising spin-glass models with a specific type of correlated bond disorder have recently been used to relate properties of spin-glass phases, such as temperature chaos and magnetic-field chaos, to ferromagnetic order. This correlated disorder has a simple physical interpretation, which underlies the sampling procedure introduced by Oshima, Arai, and Hukushima for numerical simulations: the bond configuration is obtained from the Edwards--Anderson model with the symmetric $\pm J$ distribution by drawing one of its equilibrium spin configurations and writing the model in the gauge in which this reference configuration becomes the all-up state. The construction is analogous to the Mattis model, with the roles of the ferromagnet and the spin glass reversed: whereas the Mattis model is a ferromagnet in disguise, the ferromagnetic phase of the correlated model is essentially a spin-glass phase in disguise. The frustration statistics of the correlated-disorder model are exactly those of the symmetric $\pm J$ model, and, in a generalization with an additional parameter $\gamma$, the probabilities of the gauge equivalence classes are determined by $\gamma$ alone, whereas the inverse temperature at which the reference configuration is drawn determines only the distribution within each class. The magnetization of this model is the overlap with the reference configuration. Through this correspondence, studies of its ferromagnetic phase have led to unexpected results on temperature chaos and magnetic-field chaos in the spin-glass phase of the Edwards--Anderson model. Models with correlated disorder have thus provided a new perspective on spin glasses and have proved fertile ground for highly nontrivial results on the spin-glass problem.
\end{abstract}
\date{\today}
\maketitle

\section{Introduction}

Gauge symmetry is one of the few tools that give exact results for finite-dimensional spin glasses~\cite{Nishimori1981,Nishimori2001}. Recently, a series of studies has considered the Ising model with the correlated bond distribution~\cite{Nishimori2024,Nishimori2025}
\begin{equation}
P(\tau)=\frac{1}{A}\,\frac{e^{\beta_p\sum_{\langle ij\rangle}\tau_{ij}}}{Z_\tau(\beta_p)},
\label{eq:Ptau}
\end{equation}
with
\begin{equation}
Z_\tau(\beta_p)=\sum_S e^{\beta_p\sum_{\langle ij\rangle}\tau_{ij}S_iS_j},
\label{eq:Ztau}
\end{equation}
where $\tau_{ij}=\pm1$, the sum over $S$ runs over all spin configurations, and $A=2^{N_B-N}$ is the normalization constant, with $N$ the number of sites and $N_B$ the number of bonds. The ferromagnetic phase of this model behaves very differently from that of an ordinary ferromagnet. The distribution of the magnetization is equal to the distribution of the spin overlap in the Edwards--Anderson (EA) model~\cite{Edwards1975} with the symmetric $\pm J$ distribution~\cite{Nishimori2024}, and, if the EA model exhibits magnetic-field chaos, the ferromagnetic phase becomes unstable under random fields in any dimension~\cite{Nishimori2025}, whereas an ordinary ferromagnet remains ordered under weak random fields in three and higher dimensions (Sec.~\ref{sec:gain}). Nishimori, Ohzeki, and Okuyama~\cite{Nishimori2025a} (NOO hereafter) generalized the distribution~\eqref{eq:Ptau} by introducing a parameter $\gamma$ that controls the strength of frustration and derived a logical relation between temperature chaos and reentrant phase transitions.

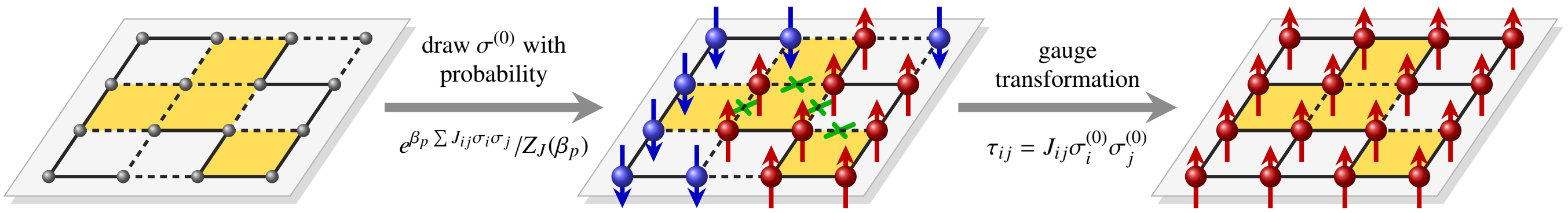
\begin{figure*}[t]
\centering
\begin{tikzpicture}[
  plate/.style={fill=black!4,draw=black!30,line width=0.4pt},
  frus/.style={fill=orange!40!yellow!65},
  bondp/.style={line width=1.1pt,black!85},
  bondm/.style={line width=1.1pt,black!85,dash pattern=on 2.2pt off 1.8pt},
  unsat/.style={green!70!black,line width=1.6pt,line cap=round},
  spin/.style={line width=1.5pt,-{Stealth[length=2.3mm,width=2.1mm]}},
  up/.style={red!75!black},down/.style={blue!75!black},
  flow/.style={line width=3pt,black!45,-{Stealth[length=3.6mm,width=3.6mm]}}]
\colorlet{upball}{red!80!black}\colorlet{downball}{blue!70!white}
\fill[black!14] (-0.437,-0.311) -- (2.827,-0.311) -- (4.198,1.712) -- (0.934,1.712) -- cycle;
\filldraw[plate] (-0.507,-0.221) -- (2.757,-0.221) -- (4.128,1.802) -- (0.864,1.802) -- cycle;
\fill[frus] (1.700,0.000) -- (2.550,0.000) -- (2.907,0.527) -- (2.057,0.527) -- cycle;
\fill[frus] (0.357,0.527) -- (1.207,0.527) -- (1.564,1.054) -- (0.714,1.054) -- cycle;
\fill[frus] (1.207,0.527) -- (2.057,0.527) -- (2.414,1.054) -- (1.564,1.054) -- cycle;
\fill[frus] (1.564,1.054) -- (2.414,1.054) -- (2.771,1.581) -- (1.921,1.581) -- cycle;
\draw[bondp] (0.000,0.000) -- (0.850,0.000);
\draw[bondp] (0.000,0.000) -- (0.357,0.527);
\draw[bondm] (0.850,0.000) -- (1.700,0.000);
\draw[bondm] (0.850,0.000) -- (1.207,0.527);
\draw[bondp] (1.700,0.000) -- (2.550,0.000);
\draw[bondp] (1.700,0.000) -- (2.057,0.527);
\draw[bondp] (2.550,0.000) -- (2.907,0.527);
\draw[bondm] (0.357,0.527) -- (1.207,0.527);
\draw[bondp] (0.357,0.527) -- (0.714,1.054);
\draw[bondp] (1.207,0.527) -- (2.057,0.527);
\draw[bondm] (1.207,0.527) -- (1.564,1.054);
\draw[bondm] (2.057,0.527) -- (2.907,0.527);
\draw[bondm] (2.057,0.527) -- (2.414,1.054);
\draw[bondp] (2.907,0.527) -- (3.264,1.054);
\draw[bondm] (0.714,1.054) -- (1.564,1.054);
\draw[bondp] (0.714,1.054) -- (1.071,1.581);
\draw[bondm] (1.564,1.054) -- (2.414,1.054);
\draw[bondm] (1.564,1.054) -- (1.921,1.581);
\draw[bondp] (2.414,1.054) -- (3.264,1.054);
\draw[bondp] (2.414,1.054) -- (2.771,1.581);
\draw[bondm] (3.264,1.054) -- (3.621,1.581);
\draw[bondp] (1.071,1.581) -- (1.921,1.581);
\draw[bondm] (1.921,1.581) -- (2.771,1.581);
\draw[bondm] (2.771,1.581) -- (3.621,1.581);
\shade[ball color=black!45] (1.071,1.581) circle (0.075);
\shade[ball color=black!45] (1.921,1.581) circle (0.075);
\shade[ball color=black!45] (2.771,1.581) circle (0.075);
\shade[ball color=black!45] (3.621,1.581) circle (0.075);
\shade[ball color=black!45] (0.714,1.054) circle (0.075);
\shade[ball color=black!45] (1.564,1.054) circle (0.075);
\shade[ball color=black!45] (2.414,1.054) circle (0.075);
\shade[ball color=black!45] (3.264,1.054) circle (0.075);
\shade[ball color=black!45] (0.357,0.527) circle (0.075);
\shade[ball color=black!45] (1.207,0.527) circle (0.075);
\shade[ball color=black!45] (2.057,0.527) circle (0.075);
\shade[ball color=black!45] (2.907,0.527) circle (0.075);
\shade[ball color=black!45] (0.000,0.000) circle (0.075);
\shade[ball color=black!45] (0.850,0.000) circle (0.075);
\shade[ball color=black!45] (1.700,0.000) circle (0.075);
\shade[ball color=black!45] (2.550,0.000) circle (0.075);
\fill[black!14] (6.113,-0.311) -- (9.377,-0.311) -- (10.748,1.712) -- (7.484,1.712) -- cycle;
\filldraw[plate] (6.043,-0.221) -- (9.307,-0.221) -- (10.678,1.802) -- (7.414,1.802) -- cycle;
\fill[frus] (8.250,0.000) -- (9.100,0.000) -- (9.457,0.527) -- (8.607,0.527) -- cycle;
\fill[frus] (6.907,0.527) -- (7.757,0.527) -- (8.114,1.054) -- (7.264,1.054) -- cycle;
\fill[frus] (7.757,0.527) -- (8.607,0.527) -- (8.964,1.054) -- (8.114,1.054) -- cycle;
\fill[frus] (8.114,1.054) -- (8.964,1.054) -- (9.321,1.581) -- (8.471,1.581) -- cycle;
\draw[unsat] (7.779,0.722) -- (8.092,0.859) (7.871,0.859) -- (8.000,0.722);
\draw[unsat] (8.875,0.458) -- (9.189,0.596) (8.968,0.596) -- (9.096,0.458);
\draw[unsat] (8.629,0.722) -- (8.942,0.859) (8.721,0.859) -- (8.850,0.722);
\draw[unsat] (8.382,0.985) -- (8.696,1.123) (8.475,1.123) -- (8.603,0.985);
\draw[bondp] (6.550,0.000) -- (7.400,0.000);
\draw[bondp] (6.550,0.000) -- (6.907,0.527);
\draw[bondm] (7.400,0.000) -- (8.250,0.000);
\draw[bondm] (7.400,0.000) -- (7.757,0.527);
\draw[bondp] (8.250,0.000) -- (9.100,0.000);
\draw[bondp] (8.250,0.000) -- (8.607,0.527);
\draw[bondp] (9.100,0.000) -- (9.457,0.527);
\draw[bondm] (6.907,0.527) -- (7.757,0.527);
\draw[bondp] (6.907,0.527) -- (7.264,1.054);
\draw[bondp] (7.757,0.527) -- (8.607,0.527);
\draw[bondm] (7.757,0.527) -- (8.114,1.054);
\draw[bondm] (8.607,0.527) -- (9.457,0.527);
\draw[bondm] (8.607,0.527) -- (8.964,1.054);
\draw[bondp] (9.457,0.527) -- (9.814,1.054);
\draw[bondm] (7.264,1.054) -- (8.114,1.054);
\draw[bondp] (7.264,1.054) -- (7.621,1.581);
\draw[bondm] (8.114,1.054) -- (8.964,1.054);
\draw[bondm] (8.114,1.054) -- (8.471,1.581);
\draw[bondp] (8.964,1.054) -- (9.814,1.054);
\draw[bondp] (8.964,1.054) -- (9.321,1.581);
\draw[bondm] (9.814,1.054) -- (10.171,1.581);
\draw[bondp] (7.621,1.581) -- (8.471,1.581);
\draw[bondm] (8.471,1.581) -- (9.321,1.581);
\draw[bondm] (9.321,1.581) -- (10.171,1.581);
\draw[spin,down] (7.621,1.941) -- (7.621,1.221);
\shade[ball color=downball] (7.621,1.581) circle (0.125);
\draw[spin,down] (8.471,1.941) -- (8.471,1.221);
\shade[ball color=downball] (8.471,1.581) circle (0.125);
\draw[spin,up] (9.321,1.221) -- (9.321,1.941);
\shade[ball color=upball] (9.321,1.581) circle (0.125);
\draw[spin,down] (10.171,1.941) -- (10.171,1.221);
\shade[ball color=downball] (10.171,1.581) circle (0.125);
\draw[spin,down] (7.264,1.414) -- (7.264,0.694);
\shade[ball color=downball] (7.264,1.054) circle (0.125);
\draw[spin,up] (8.114,0.694) -- (8.114,1.414);
\shade[ball color=upball] (8.114,1.054) circle (0.125);
\draw[spin,up] (8.964,0.694) -- (8.964,1.414);
\shade[ball color=upball] (8.964,1.054) circle (0.125);
\draw[spin,up] (9.814,0.694) -- (9.814,1.414);
\shade[ball color=upball] (9.814,1.054) circle (0.125);
\draw[spin,down] (6.907,0.887) -- (6.907,0.167);
\shade[ball color=downball] (6.907,0.527) circle (0.125);
\draw[spin,up] (7.757,0.167) -- (7.757,0.887);
\shade[ball color=upball] (7.757,0.527) circle (0.125);
\draw[spin,up] (8.607,0.167) -- (8.607,0.887);
\shade[ball color=upball] (8.607,0.527) circle (0.125);
\draw[spin,up] (9.457,0.167) -- (9.457,0.887);
\shade[ball color=upball] (9.457,0.527) circle (0.125);
\draw[spin,down] (6.550,0.360) -- (6.550,-0.360);
\shade[ball color=downball] (6.550,0.000) circle (0.125);
\draw[spin,down] (7.400,0.360) -- (7.400,-0.360);
\shade[ball color=downball] (7.400,0.000) circle (0.125);
\draw[spin,up] (8.250,-0.360) -- (8.250,0.360);
\shade[ball color=upball] (8.250,0.000) circle (0.125);
\draw[spin,up] (9.100,-0.360) -- (9.100,0.360);
\shade[ball color=upball] (9.100,0.000) circle (0.125);
\fill[black!14] (12.663,-0.311) -- (15.927,-0.311) -- (17.298,1.712) -- (14.034,1.712) -- cycle;
\filldraw[plate] (12.593,-0.221) -- (15.857,-0.221) -- (17.228,1.802) -- (13.964,1.802) -- cycle;
\fill[frus] (14.800,0.000) -- (15.650,0.000) -- (16.007,0.527) -- (15.157,0.527) -- cycle;
\fill[frus] (13.457,0.527) -- (14.307,0.527) -- (14.664,1.054) -- (13.814,1.054) -- cycle;
\fill[frus] (14.307,0.527) -- (15.157,0.527) -- (15.514,1.054) -- (14.664,1.054) -- cycle;
\fill[frus] (14.664,1.054) -- (15.514,1.054) -- (15.871,1.581) -- (15.021,1.581) -- cycle;
\draw[bondp] (13.100,0.000) -- (13.950,0.000);
\draw[bondp] (13.100,0.000) -- (13.457,0.527);
\draw[bondp] (13.950,0.000) -- (14.800,0.000);
\draw[bondp] (13.950,0.000) -- (14.307,0.527);
\draw[bondp] (14.800,0.000) -- (15.650,0.000);
\draw[bondp] (14.800,0.000) -- (15.157,0.527);
\draw[bondp] (15.650,0.000) -- (16.007,0.527);
\draw[bondp] (13.457,0.527) -- (14.307,0.527);
\draw[bondp] (13.457,0.527) -- (13.814,1.054);
\draw[bondp] (14.307,0.527) -- (15.157,0.527);
\draw[bondm] (14.307,0.527) -- (14.664,1.054);
\draw[bondm] (15.157,0.527) -- (16.007,0.527);
\draw[bondm] (15.157,0.527) -- (15.514,1.054);
\draw[bondp] (16.007,0.527) -- (16.364,1.054);
\draw[bondp] (13.814,1.054) -- (14.664,1.054);
\draw[bondp] (13.814,1.054) -- (14.171,1.581);
\draw[bondm] (14.664,1.054) -- (15.514,1.054);
\draw[bondp] (14.664,1.054) -- (15.021,1.581);
\draw[bondp] (15.514,1.054) -- (16.364,1.054);
\draw[bondp] (15.514,1.054) -- (15.871,1.581);
\draw[bondp] (16.364,1.054) -- (16.721,1.581);
\draw[bondp] (14.171,1.581) -- (15.021,1.581);
\draw[bondp] (15.021,1.581) -- (15.871,1.581);
\draw[bondp] (15.871,1.581) -- (16.721,1.581);
\draw[spin,up] (14.171,1.221) -- (14.171,1.941);
\shade[ball color=upball] (14.171,1.581) circle (0.125);
\draw[spin,up] (15.021,1.221) -- (15.021,1.941);
\shade[ball color=upball] (15.021,1.581) circle (0.125);
\draw[spin,up] (15.871,1.221) -- (15.871,1.941);
\shade[ball color=upball] (15.871,1.581) circle (0.125);
\draw[spin,up] (16.721,1.221) -- (16.721,1.941);
\shade[ball color=upball] (16.721,1.581) circle (0.125);
\draw[spin,up] (13.814,0.694) -- (13.814,1.414);
\shade[ball color=upball] (13.814,1.054) circle (0.125);
\draw[spin,up] (14.664,0.694) -- (14.664,1.414);
\shade[ball color=upball] (14.664,1.054) circle (0.125);
\draw[spin,up] (15.514,0.694) -- (15.514,1.414);
\shade[ball color=upball] (15.514,1.054) circle (0.125);
\draw[spin,up] (16.364,0.694) -- (16.364,1.414);
\shade[ball color=upball] (16.364,1.054) circle (0.125);
\draw[spin,up] (13.457,0.167) -- (13.457,0.887);
\shade[ball color=upball] (13.457,0.527) circle (0.125);
\draw[spin,up] (14.307,0.167) -- (14.307,0.887);
\shade[ball color=upball] (14.307,0.527) circle (0.125);
\draw[spin,up] (15.157,0.167) -- (15.157,0.887);
\shade[ball color=upball] (15.157,0.527) circle (0.125);
\draw[spin,up] (16.007,0.167) -- (16.007,0.887);
\shade[ball color=upball] (16.007,0.527) circle (0.125);
\draw[spin,up] (13.100,-0.360) -- (13.100,0.360);
\shade[ball color=upball] (13.100,0.000) circle (0.125);
\draw[spin,up] (13.950,-0.360) -- (13.950,0.360);
\shade[ball color=upball] (13.950,0.000) circle (0.125);
\draw[spin,up] (14.800,-0.360) -- (14.800,0.360);
\shade[ball color=upball] (14.800,0.000) circle (0.125);
\draw[spin,up] (15.650,-0.360) -- (15.650,0.360);
\shade[ball color=upball] (15.650,0.000) circle (0.125);
\node[font=\small] at (1.560,-1.05) {(a) Step (i): bonds $J_{ij}=\pm1$};
\node[font=\small] at (8.111,-1.05) {(b) Step (ii): reference configuration $\sigma^{(0)}$};
\node[font=\small] at (14.660,-1.05) {(c) Step (iii): bonds $\tau_{ij}$};
\draw[flow] (3.835,0.790) -- (6.335,0.790);
\node[font=\footnotesize,above,align=center] at (5.085,0.910) {draw $\sigma^{(0)}$ with\\probability};
\node[font=\footnotesize,below] at (5.085,0.640) {$e^{\beta_p\sum J_{ij}\sigma_i\sigma_j}/Z_J(\beta_p)$};
\draw[flow] (10.386,0.790) -- (12.886,0.790);
\node[font=\footnotesize,above,align=center] at (11.636,0.910) {gauge\\transformation};
\node[font=\footnotesize,below] at (11.636,0.640) {$\tau_{ij}=J_{ij}\sigma^{(0)}_i\sigma^{(0)}_j$};
\end{tikzpicture}
\caption{\label{fig:procedure}Generation of the correlated disorder (Sec.~\ref{sec:procedure}), illustrated for a $4\times4$ lattice with free boundary conditions. Solid and dashed lines denote bonds $+1$ and $-1$, respectively, and colored plaquettes are frustrated. (a) Step (i): the bonds $J_{ij}$ are drawn independently, $\pm1$ with probability $1/2$ each. (b) Step (ii): a reference configuration $\sigma^{(0)}$ (up spins in red, down spins in blue) is drawn from the Boltzmann distribution for the bonds $J$ at inverse temperature $\beta_p$; green crosses mark the bonds that are not satisfied by $\sigma^{(0)}$. (c) Step (iii): after the gauge transformation, the reference configuration is the all-up state, and the bonds $\tau_{ij}$ are $-1$ exactly on the bonds marked in (b). In (b), the bonds are given and $\sigma^{(0)}$ is random, drawn from the equilibrium distribution at inverse temperature $\beta_p$; in (c), the all-up state is fixed and the bonds $\tau_{ij}$ are random, distributed so that the all-up state has the energy and the pattern of unsatisfied bonds of an equilibrium configuration at $\beta_p$. The frustrated plaquettes are the same in the three panels. The reference configuration in (b), and hence the all-up state in (c), was drawn at $\beta_p=1$ and is not a ground state; it has four unsatisfied bonds.}
\end{figure*}

Although the distribution~\eqref{eq:Ptau} appears to have complicated correlations, it is in fact the distribution of the bonds of the EA model with the symmetric distribution, gauge transformed by an equilibrium spin configuration of the same system. This is closely analogous to the Mattis model~\cite{Mattis1976}, which looks random but is obtained from a ferromagnet by a gauge transformation with a random configuration unrelated to the bonds. The roles of the ferromagnet and the spin glass are, however, reversed: whereas the Mattis model is a ferromagnet in disguise, the ferromagnetic phase of the correlated model is a spin-glass phase in disguise (Sec.~\ref{sec:identity}). Once this interpretation is understood, the results of Refs.~\cite{Nishimori2024,Nishimori2025} can be read off almost directly, and the extension by NOO can be understood from the same viewpoint.

Oshima, Arai, and Hukushima~\cite{Oshima2026} (OAH hereafter) showed that bonds following the distribution of NOO, which are difficult to generate directly because the bonds are correlated through the partition function~\cite{Nishimori2024}, can be generated exactly by a gauge transformation using an equilibrium configuration of the EA model, and used this procedure in numerical simulations in two dimensions. They also pointed out that the magnetization of the correlated model is then the overlap between equilibrium configurations at two, generally different, temperatures, which relates it to temperature chaos.

This paper makes explicit the physical content of the construction of OAH: an equilibrium reference configuration is embedded by choosing the gauge in which it becomes the all-up state. We organize this construction in terms of gauge equivalence classes, which separates the roles of the frustration and of the choice of gauge, and use the correspondence between the magnetization and the overlap to explain the consequences of temperature chaos and magnetic-field chaos for the ferromagnetic order of the correlated model.

Most of the results discussed below are known or follow directly from known results; the aim of this paper is to present them in a unified picture, not to claim new physical results.

This paper is organized as follows. Sections~\ref{sec:construction}--\ref{sec:identity} contain exact statements for finite systems: the construction of the correlated disorder by embedding an equilibrium spin configuration (Sec.~\ref{sec:construction}), its decomposition into gauge equivalence classes (Sec.~\ref{sec:class}), and the identification of the magnetization with an overlap (Sec.~\ref{sec:identity}), from which the consequences of assuming temperature chaos in the EA model are drawn (Sec.~\ref{sec:NL}). Section~\ref{sec:field} turns to random fields, with the consequences of assuming magnetic-field chaos (Sec.~\ref{sec:overlap-field}) and the intuition from the droplet picture (Secs.~\ref{sec:wall} and \ref{sec:gain}). Section~\ref{sec:IsingXY} compares the range covered by rigorous proofs for Ising and $XY$ spins, Sec.~\ref{sec:discussion} discusses the uses of the correspondence, and Sec.~\ref{sec:summary} summarizes the paper. The gauge correspondence has the same structure for the $XY$ gauge glass.

\section{Construction of the correlated disorder}\label{sec:construction}

\subsection{Generation procedure}\label{sec:procedure}

We consider a finite connected lattice with free or periodic boundary conditions. The distribution~\eqref{eq:Ptau} is exactly the distribution of the bonds generated by the following procedure.
\begin{romanlist}
\item Draw the bonds $J_{ij}$ of the symmetric $\pm J$ model, $\pm1$ with probability $1/2$ each, independently for each bond.
\item Draw one spin configuration $\sigma^{(0)}$ from the Boltzmann distribution of the EA model with the bonds $J_{ij}$ drawn in step (i), at inverse temperature $\beta_p$ and zero field. We call it the \textit{reference configuration}.
\item Apply the gauge transformation $\tau_{ij}=J_{ij}\sigma^{(0)}_i\sigma^{(0)}_j$.
\end{romanlist}
Figure~\ref{fig:procedure} illustrates the procedure for a small lattice. This procedure is the case $\gamma=0$ of the procedure given explicitly by OAH for the general form with $\gamma$ (Sec.~\ref{sec:gamma}) [Eq.~(9) of Ref.~\cite{Oshima2026}].

\subsection{Derivation of the distribution}\label{sec:derivation}

In steps (i) and (ii), the probability of obtaining the pair of bonds $J$ and reference configuration $\sigma^{(0)}$ is
\begin{equation}
P(J,\sigma^{(0)})=2^{-N_B}\times\frac{e^{\beta_p\sum J_{ij}\sigma^{(0)}_i\sigma^{(0)}_j}}{Z_J(\beta_p)} .
\end{equation}
The first factor is the uniform distribution of $J$, and the second factor is the Boltzmann distribution of $\sigma^{(0)}$ for given $J$.

The probability of a given $\tau$ is the sum over all pairs $(J,\sigma^{(0)})$ that give this $\tau$. For given $\sigma^{(0)}$, the bonds are uniquely determined as $J_{ij}=\tau_{ij}\sigma^{(0)}_i\sigma^{(0)}_j$, since $(\sigma^{(0)}_i)^2=1$. Hence
\begin{equation}
P(\tau)=\sum_{\sigma^{(0)}}2^{-N_B}\left.\frac{e^{\beta_p\sum J_{ij}\sigma^{(0)}_i\sigma^{(0)}_j}}{Z_J(\beta_p)}\right|_{J_{ij}=\tau_{ij}\sigma^{(0)}_i\sigma^{(0)}_j}.
\end{equation}
Two facts can now be used.
\begin{romanlist}
\item Numerator: $J_{ij}\sigma^{(0)}_i\sigma^{(0)}_j=\tau_{ij}(\sigma^{(0)}_i)^2(\sigma^{(0)}_j)^2=\tau_{ij}$, so that the numerator becomes $e^{\beta_p\sum\tau_{ij}}$.
\item Denominator: since the sum over spins runs over all configurations, the change of summation variables $S_i\to S_i\sigma^{(0)}_i$ gives $Z_J=Z_\tau$ (gauge invariance of the partition function).
\end{romanlist}
The summand is therefore independent of $\sigma^{(0)}$, and the sum only gives a factor $2^N$:
\begin{equation}
P(\tau)=2^{N-N_B}\,\frac{e^{\beta_p\sum\tau_{ij}}}{Z_\tau(\beta_p)} .
\end{equation}
This is Eq.~\eqref{eq:Ptau} itself, including the normalization constant $A=2^{N_B-N}$. This result is an identity for finite systems, and it does not require the existence of a spin-glass phase or of chaos. If the boundary spins are fixed, the gauge transformation also changes the boundary spins, and the derivation above has to be reexamined.

The bond $\tau_{ij}=J_{ij}\sigma^{(0)}_i\sigma^{(0)}_j$ is $+1$ if the bond $ij$ is satisfied by the reference configuration $\sigma^{(0)}$ and $-1$ otherwise. In other words, $\tau$ is \textit{the map of which bonds are satisfied by the reference configuration} [Figs.~\ref{fig:procedure}(b) and \ref{fig:procedure}(c)]. The original bonds $J_{ij}$ are mutually independent, but $\sigma^{(0)}$ is a single configuration chosen by looking at all the bonds at once, and this map is therefore, in general, not independent from bond to bond. (On a lattice without loops, such as a tree, $Z_\tau(\beta_p)$ does not depend on $\tau$, and the bonds remain independent.)

\subsection{Limit \texorpdfstring{$\beta_p\to\infty$}{beta\_p to infinity}}\label{sec:lowT}

Here we take the limit $\beta_p\to\infty$ for a finite system. In this limit, the correlated distribution concentrates on the bond configurations for which the all-up state (the pure ferromagnetic state of Ref.~\cite{Nishimori2024}) is one of the ground states~\cite{Nishimori2024}. We explain why the all-up state plays this special role.

We denote the energy of a configuration $S$ of the correlated model by $E_\tau(S)=-\sum\tau_{ij}S_iS_j$, and the all-up state ($S_i=1$ for all $i$) by $\bm{1}$, so that $E_\tau(\bm{1})=-\sum\tau_{ij}$. The factor in Eq.~\eqref{eq:Ptau} is then
\begin{equation}
\frac{e^{\beta_p\sum\tau_{ij}}}{Z_\tau(\beta_p)}=\frac{e^{-\beta_pE_\tau(\bm{1})}}{Z_\tau(\beta_p)} ,
\label{eq:allupprob}
\end{equation}
which is the Boltzmann probability of the all-up state in the system with bonds $\tau$ at inverse temperature $\beta_p$. The correlated distribution~\eqref{eq:Ptau} is thus proportional to this probability, $P(\tau)=2^{N-N_B}e^{-\beta_pE_\tau(\bm{1})}/Z_\tau(\beta_p)$: it favors bond configurations in which the all-up state is likely to be found in equilibrium at inverse temperature $\beta_p$. In the embedding picture, this reflects the fact that the all-up state is the image of the reference configuration $\sigma^{(0)}$, which is drawn from the Boltzmann distribution at $\beta_p$. Indeed, with the energy $E_J(S')=-\sum J_{ij}S'_iS'_j$ of a configuration $S'$ of the original EA model and $S'_i=S_i\sigma^{(0)}_i$, we have $E_\tau(S)=E_J(S')$ for every $S$, and the all-up state corresponds to $S'=\sigma^{(0)}$.

For a finite system, the Boltzmann probability of a configuration tends, as $\beta_p\to\infty$, to $1/\mathcal{N}_\tau$ if the configuration is one of the $\mathcal{N}_\tau$ ground states, and to zero otherwise. Hence $P(\tau)$ tends to $2^{N-N_B}/\mathcal{N}_\tau$ if the all-up state is a ground state of the system with bonds $\tau$, and to zero otherwise. In the embedding picture, $\sigma^{(0)}$ is then a ground state of the system with bonds $J$, and $E_\tau(\bm{1})=E_J(\sigma^{(0)})=\min_{S'}E_J(S')=\min_SE_\tau(S)$. Even apart from its global reversal, the all-up state is in general not the only ground state, because the ground states of the $\pm J$ model are degenerate. For example, if two consecutive horizontal bonds on the square lattice are antiferromagnetic and all other bonds are ferromagnetic, the spin between these two bonds can point up or down at the same energy (footnote~1 of Ref.~\cite{Nishimori2024}).

At finite $\beta_p$, the pair of the bonds $\tau$ and the all-up state is the gauge image of the pair of $J$ and $\sigma^{(0)}$. In the description with the bonds $J$, the reference configuration $\sigma^{(0)}$ is random: it is drawn from the equilibrium distribution and is, in this sense, a typical equilibrium configuration. In the description with the bonds $\tau$, the all-up state is a fixed configuration, and it is the bonds that are random: they are drawn with a probability proportional to the Boltzmann probability~\eqref{eq:allupprob} of the all-up state. The bonds $\tau$ are thus chosen so that the all-up state has the properties of an equilibrium configuration at inverse temperature $\beta_p$. More precisely, quantities that are invariant under the gauge transformation, such as the energy, the number of unsatisfied bonds, and the number of spins that are unstable against a single flip, take the same values for the all-up state with the bonds $\tau$ as for $\sigma^{(0)}$ with the bonds $J$, and therefore have the statistics of an equilibrium configuration. An analogous structure appears in the quiet planting of a solution in random constraint satisfaction problems by Krzakala and Zdeborov\'a~\cite{Krzakala2009,Zdeborova2016}; in the present case, the property that the frustration statistics remain those of the original model holds exactly in finite systems (Sec.~\ref{sec:frustration}).

\subsection{Generalization with \texorpdfstring{$\gamma$}{gamma} (NOO)}\label{sec:gamma}

NOO~\cite{Nishimori2025a} generalized Eq.~\eqref{eq:Ptau} to the distribution
\begin{equation}
P_\gamma(\tau)=\frac{1}{A_\gamma}\,Z_\tau(\gamma)\,\frac{e^{\beta_p\sum\tau_{ij}}}{Z_\tau(\beta_p)},\qquad
A_\gamma=(2\cosh\gamma)^{N_B}.
\label{eq:Pgamma}
\end{equation}
As in Eq.~\eqref{eq:Ptau}, the dependence on $\beta_p$ is not written explicitly. For $\gamma=0$, we have $Z_\tau(0)=2^N$, and Eq.~\eqref{eq:Ptau} is recovered. For $\gamma=\beta_p$, the two partition functions cancel, and the distribution becomes that of independent bonds,
\begin{equation}
P_{\rm EA}(\tau;\beta_p)=\prod_{\langle ij\rangle}\frac{e^{\beta_p\tau_{ij}}}{2\cosh\beta_p},
\label{eq:PEA}
\end{equation}
namely the EA model in which the probability $p$ of a ferromagnetic bond is given by $e^{-2\beta_p}=(1-p)/p$; $\beta_p$ is the inverse temperature on its Nishimori line. For $\beta_p=0$, $P_\gamma\propto Z_\tau(\gamma)$. This is a gauge-invariant distribution~\cite{Nishimori2025a}; it has been proposed as the distribution generated by quantum circuits with measurements~\cite{Zhu2023} and realized experimentally on a quantum computer~\cite{Chen2025}. The model of Kitatani~\cite{Kitatani1992} corresponds to $\gamma=K_p$ and $\beta_p=K_p+a$ in his notation. For $\gamma\to\infty$, only the unfrustrated configurations $\tau_{ij}=\xi_i\xi_j$ of the Mattis type remain~\cite{Nishimori2025a}.

The generation procedure is obtained by replacing step (i) of Sec.~\ref{sec:procedure} by drawing $J$ from the distribution $P_{\rm EA}(J;\gamma)$ of the EA model~\cite{Oshima2026}. Indeed, the probability of the pair $(J,\sigma^{(0)})$ is then
\begin{equation}
P(J,\sigma^{(0)})=\frac{e^{\gamma\sum J_{ij}}}{(2\cosh\gamma)^{N_B}}\times\frac{e^{\beta_p\sum J_{ij}\sigma^{(0)}_i\sigma^{(0)}_j}}{Z_J(\beta_p)} .
\end{equation}
As in Sec.~\ref{sec:derivation}, the probability of a given $\tau$ is the sum of this probability over all pairs $(J,\sigma^{(0)})$ that give this $\tau$. For fixed $\tau$, the bonds $J_{ij}=\tau_{ij}\sigma^{(0)}_i\sigma^{(0)}_j$ are uniquely determined by $\sigma^{(0)}$, so that the sum is over $\sigma^{(0)}$. On substitution, the numerator of the first factor becomes $e^{\gamma\sum\tau_{ij}\sigma^{(0)}_i\sigma^{(0)}_j}$, and the second factor becomes $e^{\beta_p\sum\tau_{ij}}/Z_\tau(\beta_p)$. The second factor does not depend on $\sigma^{(0)}$, and the sum turns the numerator of the first factor into $Z_\tau(\gamma)$, which gives Eq.~\eqref{eq:Pgamma}. Setting $\gamma=0$, we recover the calculation of Sec.~\ref{sec:derivation}. Figure~\ref{fig:procedure} applies to $\gamma>0$ without change, except that each bond in panel (a) is $+1$ with probability $e^{\gamma}/(2\cosh\gamma)$, so that dashed bonds and frustrated plaquettes are typically fewer.

The bonds $J$ chosen in step (i) determine the configuration of frustration, and steps (ii) and (iii) determine the gauge under the same frustration. OAH describe this as follows: the configuration of frustration is determined by $J$, and the specific bond configuration is determined by the gauge $\sigma$~\cite{Oshima2026}. We show how this division of roles holds exactly in Sec.~\ref{sec:class-gamma}.

\section{Decomposition into gauge equivalence classes}\label{sec:class}

\subsection{Gauge equivalence classes of bond configurations}

\begin{figure}[t]
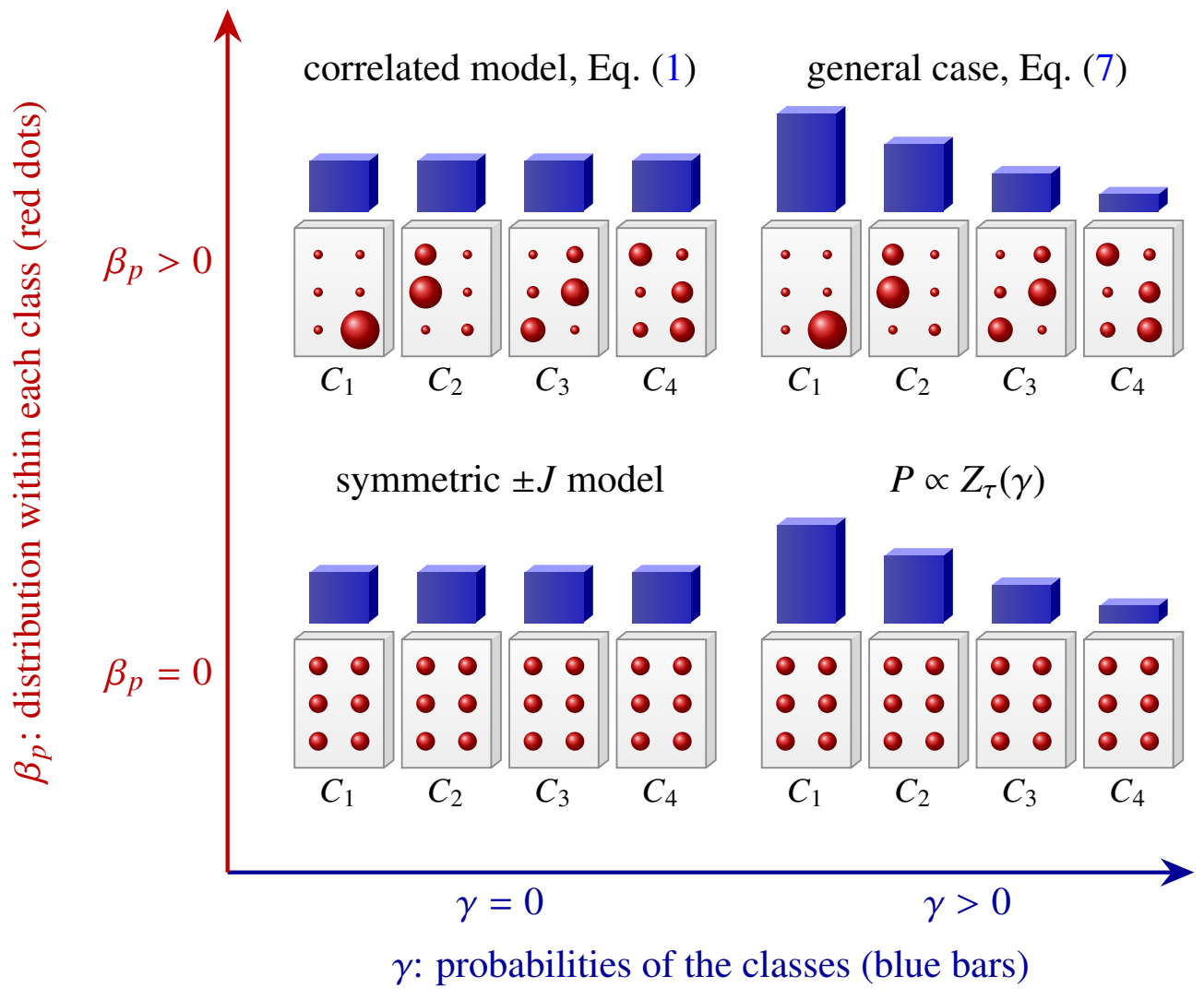

\centering

\caption{\label{fig:classes}Schematic picture of the decomposition of the bond distribution into gauge equivalence classes (Sec.~\ref{sec:class}). Each gray box is an equivalence class $C_k$; only four classes, with six of the $2^{N-1}$ bond configurations of each, are drawn, arranged schematically from weak ($C_1$) to strong ($C_4$) frustration. The height of a blue bar represents the probability that the class is chosen, $P_\gamma(C)$ of Eq.~\eqref{eq:PCgamma}. Each red dot is one bond configuration $\tau$ in the class, and its area represents the conditional probability $P(\tau|C)$ of Eq.~\eqref{eq:PtauC} that this configuration is chosen within the class. The probability of a bond configuration is the product of the two. The four panels are arranged by $\gamma$ (horizontal) and $\beta_p$ (vertical). For $\gamma=0$, all classes are equally probable; for $\gamma>0$, the bars are proportional to $Z_C(\gamma)$, and classes with weaker frustration are typically more probable. For $\beta_p=0$, all configurations in a class are equally probable; for $\beta_p>0$, the dots have the Boltzmann weights $e^{\beta_p\sum\tau_{ij}}$. A change of $\gamma$ changes only the bars, and a change of $\beta_p$ changes only the dots. The general case reduces to the EA model for $\gamma=\beta_p$.}
\end{figure}

The set of bond configurations obtained from a bond configuration $\tau$ by the gauge transformations
\begin{equation}
\tau_{ij}\ \to\ \tau^{\sigma}_{ij}=\tau_{ij}\sigma_i\sigma_j
\end{equation}
for all $\sigma\in\{\pm1\}^N$ is an \textit{equivalence class} (gauge equivalence class) of the equivalence relation ``related to each other by a gauge transformation.'' It corresponds to what is called a gauge orbit in lattice gauge theory. The set of all bond configurations is divided into mutually disjoint equivalence classes.

Configurations in the same equivalence class have the same frustration (as in the three panels of Fig.~\ref{fig:procedure}), and, conversely, configurations with the same frustration belong to the same equivalence class. Here, frustration means the products of bonds along closed loops; for periodic boundary conditions, these include the products along the loops that wind around the system, in addition to the plaquettes.

The configurations $\sigma$ and $-\sigma$ give the same $\tau^\sigma$, and no other pair of spin configurations gives the same bond configuration. Every equivalence class therefore contains the same number $2^{N-1}$ of configurations, and the number of equivalence classes is $2^{N_B}/2^{N-1}$.

\subsection{Probability of an equivalence class and probability within a class (\texorpdfstring{$\gamma=0$}{gamma=0})}\label{sec:class-prob}

In this and the next subsection, we treat the distribution~\eqref{eq:Ptau}, i.e., the case $\gamma=0$ of the generalization of NOO. The case with $\gamma$ is treated in Sec.~\ref{sec:class-gamma}.

We write $P(\tau)=P(C)\,P(\tau|C)$, where $P(C)$ is the probability that the equivalence class $C$ is chosen and $P(\tau|C)$ is the conditional probability that the configuration $\tau$ is chosen within $C$. Figure~\ref{fig:classes} gives a schematic picture of this decomposition.

\paragraph{All equivalence classes have the same probability.}
The probability of an equivalence class $C$ is the sum of $P(\tau)$ over all configurations in $C$. Writing the configurations in $C$ as $\tau^\sigma$ with a fixed $\tau\in C$ and noting that each configuration appears twice in the sum over $\sigma$, we have
\begin{align}
P(C)&=\sum_{\tau'\in C}P(\tau')=\frac12\sum_\sigma\frac1A\,\frac{e^{\beta_p\sum\tau_{ij}\sigma_i\sigma_j}}{Z_{\tau^\sigma}(\beta_p)}
\nonumber\\
&=\frac{1}{2A}=\frac{2^{N-1}}{2^{N_B}} ,
\label{eq:PC}
\end{align}
which does not depend on the equivalence class. In other words, \textit{all equivalence classes have the same probability}. In the symmetric $\pm J$ model, each configuration has probability $2^{-N_B}$, and the probability of an equivalence class is $2^{N-1}\times2^{-N_B}$, exactly the same. Hence the frustration statistics, and more generally the disorder averages of gauge-invariant quantities at zero field, are the same for the correlated model and for the symmetric $\pm J$ model.

\paragraph{The probability within a class is proportional to a Boltzmann weight.}
In contrast, the configurations within an equivalence class do not have the same probability. The conditional probability is
\begin{equation}
P(\tau'|C)=\frac{e^{\beta_p\sum\tau'_{ij}}}{\sum_{\tau''\in C}e^{\beta_p\sum\tau''_{ij}}} .
\label{eq:PtauC}
\end{equation}
If $\tau'=\tau^\sigma$, then $e^{\beta_p\sum\tau'_{ij}}=e^{\beta_p\sum\tau_{ij}\sigma_i\sigma_j}$ is the Boltzmann weight of the spin configuration $\sigma$ for the bonds $\tau$. Choosing a configuration within an equivalence class is therefore the same as choosing a spin configuration from the Boltzmann distribution of the system with these bonds. In the symmetric $\pm J$ model, the probability within an equivalence class is uniform.

The procedure of Sec.~\ref{sec:procedure} is precisely this two-stage choice. Drawing $J$ in step (i) chooses the equivalence class to which $J$ belongs, with equal probabilities. Choosing $\sigma^{(0)}$ and setting $\tau=J^{\sigma^{(0)}}$ in steps (ii) and (iii) amounts to choosing a configuration within this class with probability $P(\tau|C)\propto e^{\beta_p\sum\tau_{ij}}$.

\subsection{The frustration is that of the symmetric \texorpdfstring{$\pm J$}{+-J} model (\texorpdfstring{$\gamma=0$}{gamma=0})}\label{sec:frustration}

It follows that the correlated model differs from the symmetric $\pm J$ model only in which configuration is taken within each equivalence class, i.e., only in the choice of gauge. The frustration statistics are exactly the same as those of the symmetric $\pm J$ model with $p=1/2$.

In contrast, a pure ferromagnet has no frustration at all, and an EA model with an independent distribution biased toward ferromagnetic bonds ($p$ close to 1) has little frustration. The frustration statistics are gauge invariant, and in this respect the correlated model differs essentially from these models and is the same as the symmetric $\pm J$ model. The average of the bonds and the usual magnetization, on the other hand, depend on the gauge. For instance, a pure ferromagnet can also be transformed into the Mattis-type bonds $\tau_{ij}=\xi_i\xi_j$ by a gauge transformation. The equality of the frustration statistics does not determine the magnetization or the interface tension in a particular gauge. The ``ferromagnetism'' of the correlated model is the spin-glass order viewed in the gauge in which the reference configuration looks like the all-up state.

\subsection{\texorpdfstring{$\gamma$}{gamma} determines the frustration and \texorpdfstring{$\beta_p$}{beta\_p} the gauge}\label{sec:class-gamma}

The calculation of Sec.~\ref{sec:class-prob} extends directly to the distribution~\eqref{eq:Pgamma} of NOO. Both $Z_\tau(\gamma)$ and $Z_\tau(\beta_p)$ are gauge invariant and take common values within an equivalence class $C$. Denoting them by $Z_C(\gamma)$ and $Z_C(\beta_p)$, we have
\begin{align}
P_\gamma(C)&=\sum_{\tau'\in C}P_\gamma(\tau')
=\frac{Z_C(\gamma)}{A_\gamma\,Z_C(\beta_p)}\sum_{\tau'\in C}e^{\beta_p\sum\tau'_{ij}}
\nonumber\\
&=\frac{Z_C(\gamma)}{2A_\gamma},
\label{eq:PCgamma}
\end{align}
and the probability $P_\gamma(\tau'|C)$ within an equivalence class is exactly the same as Eq.~\eqref{eq:PtauC}. That is, the probability of an equivalence class does not depend on $\beta_p$, and the probability within a class does not depend on $\gamma$ (Fig.~\ref{fig:classes}).

\begin{figure}[t]
\centering
\begin{tikzpicture}[every node/.style={inner sep=1.5pt}]
\colorlet{gcol}{blue!60!black}\colorlet{bcol}{red!75!black}
\fill[black!5.4] (2.100,0.720) -- (2.260,0.845) -- (2.260,3.345) -- (2.100,3.220) -- cycle;
\fill[black!3.1] (0.000,3.220) -- (0.160,3.345) -- (2.260,3.345) -- (2.100,3.220) -- cycle;
\shade[top color=black!0.7,bottom color=black!2.4] (0.000,0.720) rectangle (2.100,3.220);
\draw[black!20.4,line width=0.4pt] (0.000,0.720) -- (2.100,0.720) -- (2.260,0.845) -- (2.260,3.345) -- (0.160,3.345) -- (0.000,3.220) -- cycle (2.100,0.720) -- (2.100,3.220) -- (0.000,3.220);
\fill[black!9.0] (2.400,0.480) -- (2.560,0.605) -- (2.560,3.105) -- (2.400,2.980) -- cycle;
\fill[black!5.0] (0.300,2.980) -- (0.460,3.105) -- (2.560,3.105) -- (2.400,2.980) -- cycle;
\shade[top color=black!1.1,bottom color=black!3.9] (0.300,0.480) rectangle (2.400,2.980);
\draw[black!33.6,line width=0.4pt] (0.300,0.480) -- (2.400,0.480) -- (2.560,0.605) -- (2.560,3.105) -- (0.460,3.105) -- (0.300,2.980) -- cycle (2.400,0.480) -- (2.400,2.980) -- (0.300,2.980);
\fill[black!12.5] (2.700,0.240) -- (2.860,0.365) -- (2.860,2.865) -- (2.700,2.740) -- cycle;
\fill[black!7.0] (0.600,2.740) -- (0.760,2.865) -- (2.860,2.865) -- (2.700,2.740) -- cycle;
\shade[top color=black!1.6,bottom color=black!5.5] (0.600,0.240) rectangle (2.700,2.740);
\draw[black!46.8,line width=0.4pt] (0.600,0.240) -- (2.700,0.240) -- (2.860,0.365) -- (2.860,2.865) -- (0.760,2.865) -- (0.600,2.740) -- cycle (2.700,0.240) -- (2.700,2.740) -- (0.600,2.740);
\fill[black!16.0] (3.000,0.000) -- (3.160,0.125) -- (3.160,2.625) -- (3.000,2.500) -- cycle;
\fill[black!9.0] (0.900,2.500) -- (1.060,2.625) -- (3.160,2.625) -- (3.000,2.500) -- cycle;
\shade[top color=black!2.0,bottom color=black!7.0] (0.900,0.000) rectangle (3.000,2.500);
\draw[black!60,line width=0.4pt] (0.900,0.000) -- (3.000,0.000) -- (3.160,0.125) -- (3.160,2.625) -- (1.060,2.625) -- (0.900,2.500) -- cycle (3.000,0.000) -- (3.000,2.500) -- (0.900,2.500);
\fill[black!16.0] (6.600,0.000) -- (6.760,0.125) -- (6.760,2.625) -- (6.600,2.500) -- cycle;
\fill[black!9.0] (4.500,2.500) -- (4.660,2.625) -- (6.760,2.625) -- (6.600,2.500) -- cycle;
\shade[top color=black!2.0,bottom color=black!7.0] (4.500,0.000) rectangle (6.600,2.500);
\draw[black!60,line width=0.4pt] (4.500,0.000) -- (6.600,0.000) -- (6.760,0.125) -- (6.760,2.625) -- (4.660,2.625) -- (4.500,2.500) -- cycle (6.600,0.000) -- (6.600,2.500) -- (4.500,2.500);
\node[font=\Large] at (1.950,1.250) {$C_i$};
\node[font=\normalsize,anchor=north west] at (4.580,2.440) {$C_i$};
\shade[ball color=red!85!black] (5.067,1.650) circle (0.05);
\shade[ball color=red!85!black] (5.655,1.650) circle (0.16);
\shade[ball color=red!85!black] (6.180,1.650) circle (0.06);
\shade[ball color=red!85!black] (5.067,1.050) circle (0.08);
\shade[ball color=red!85!black] (5.655,1.050) circle (0.05);
\shade[ball color=red!85!black] (6.180,1.050) circle (0.12);
\shade[ball color=red!85!black] (5.067,0.450) circle (0.11);
\shade[ball color=red!85!black] (5.655,0.450) circle (0.06);
\shade[ball color=red!85!black] (6.180,0.450) circle (0.07);
\draw[black!75,line width=0.9pt] (5.655,1.650) circle (0.260);
\node[font=\normalsize,anchor=south west] at (5.825,1.780) {$\tau$};
\draw[line width=2.2pt,black!45,-{Stealth[length=3mm,width=3mm]}] (3.360,1.250) -- (4.250,1.250);
\node[font=\Large,gcol,anchor=base] at (1.950,-0.6) {$\gamma$};
\node[font=\Large,bcol,anchor=base] at (5.550,-0.6) {$\beta_p$};
\end{tikzpicture}
\caption{\label{fig:roles}Roles of the two parameters of the distribution~\eqref{eq:Pgamma}. The parameter $\gamma$ determines the probability $P_\gamma(C_i)\propto Z_{C_i}(\gamma)$ with which an equivalence class $C_i$ is chosen among all classes (left; the other classes are drawn behind). The parameter $\beta_p$ determines the probability $P(\tau|C_i)\propto e^{\beta_p\sum\tau_{ij}}$ with which a bond configuration $\tau$ is chosen within the class, i.e., the choice of gauge (right). Each red dot is a configuration in the class, and its area represents this probability; the circled one is the configuration chosen.}
\end{figure}
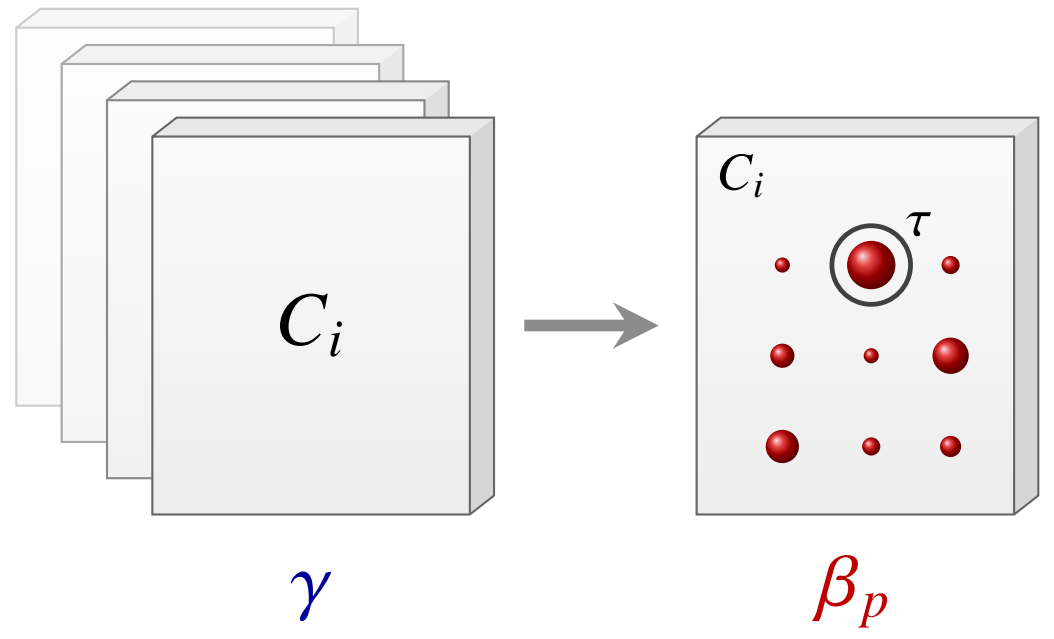

The probability~\eqref{eq:PCgamma} of an equivalence class is equal to that for the EA model $P_{\rm EA}(\tau;\gamma)$. Indeed, the same decomposition of the distribution~\eqref{eq:PEA} with $\beta_p$ replaced by $\gamma$ gives $\sum_{\tau\in C}e^{\gamma\sum\tau_{ij}}/(2\cosh\gamma)^{N_B}=Z_C(\gamma)/[2(2\cosh\gamma)^{N_B}]$. The roles of the two parameters are therefore separated in the following sense.
\begin{romanlist}
\item The parameter $\gamma$ determines only the frustration statistics, which are those of the EA model with the probability of ferromagnetic bonds corresponding to $\gamma$. We denote this EA model by EA($\gamma$).
\item The parameter $\beta_p$ determines only the choice of gauge under the same frustration. For $\beta_p=0$, the gauge is chosen uniformly at random; for $\beta_p>0$, it is chosen with the Boltzmann weight of Eq.~\eqref{eq:PtauC}, that is, by drawing one configuration $\sigma^{(0)}$ from the equilibrium distribution of EA($\gamma$) at inverse temperature $\beta_p$ and taking the gauge in which $\sigma^{(0)}$ becomes the all-up state [steps (ii) and (iii) of Sec.~\ref{sec:procedure}].
\end{romanlist}
Figure~\ref{fig:roles} illustrates this two-stage choice.
This separation refers to the marginal distribution of the equivalence classes and to the conditional distribution within a class. It does not mean that the class and the configuration within it are statistically independent random variables; the conditional distribution is defined on each class and depends on that class.
This is a restatement, in terms of distributions, of the result of NOO themselves that disorder averages of gauge-invariant quantities do not depend on $\beta_p$ and are equal to those of EA($\gamma$)~\cite{Nishimori2025a}. For example, the quenched free energy at zero field, i.e., the average of $-\beta^{-1}\ln Z_\tau(\beta)$ over the distribution~\eqref{eq:Pgamma}, is that of EA($\gamma$) for any $\beta_p$: a change of $\beta_p$ at fixed $\gamma$ leaves it unchanged and affects only gauge-dependent quantities such as the magnetization.

It is explained in Sec.~II~A of Ref.~\cite{Nishimori2024} and in Sec.~II~B of NOO that the factor $1/Z_\tau(\beta_p)$ increases the weight of strongly frustrated configurations relative to the EA model, and in the latter that the numerator $Z_\tau(\gamma)$ increases the weight of weakly frustrated configurations, partly compensating for it (OAH give a similar explanation). According to the above decomposition, $e^{\beta_p\sum\tau_{ij}}/Z_\tau(\beta_p)$ is a factor for the choice of gauge that gives the same total weight to every equivalence class, $\sum_{\tau'\in C}e^{\beta_p\sum\tau'_{ij}}/Z_C(\beta_p)=1/2$ by Eq.~\eqref{eq:PC}, so that it has no effect at all on the frustration statistics. The statement that $1/Z_\tau(\beta_p)$ enhances strong frustration applies only in comparison with the EA model $P_{\rm EA}(\tau;\beta_p)$, whose probability of an equivalence class is proportional to $Z_C(\beta_p)$. The frustration statistics are determined by $Z_\tau(\gamma)$ alone.

\section{The correlated model as an embedded spin glass}\label{sec:identity}

\subsection{Two gauges}\label{sec:twogauges}

We consider the correlated model at inverse temperature $\beta$ and zero field, with the Boltzmann weight $e^{\beta\sum\tau_{ij}S_iS_j}$.

In preparation for Sec.~\ref{sec:overlap}, we recall the correspondence used in Sec.~\ref{sec:lowT}. For $\tau_{ij}=J_{ij}\sigma^{(0)}_i\sigma^{(0)}_j$, associating with each spin configuration $S$ of the correlated model the configuration $S'_i=S_i\sigma^{(0)}_i$ of the original EA model, we have
\begin{equation}
\sum\tau_{ij}S_iS_j=\sum J_{ij}S'_iS'_j .
\end{equation}
The same system can thus be expressed in two ways, in terms of the bonds $\tau$ and the configuration $S$ (the gauge of the correlated model), and in terms of the bonds $J$ and the configuration $S'$ (the gauge of the original EA model). The all-up state $S_i=1$ of the correlated model corresponds to the reference configuration $\sigma^{(0)}$ in the gauge of the original EA model. The correlated model is therefore the EA model in which one configuration drawn from its equilibrium distribution is embedded and written in the gauge in which this configuration looks like the all-up state.

The all-up state is the image of the reference configuration at any temperature and system size. Whether the correlated model has a macroscopic magnetization is a separate question, which is answered by the overlap statistics of the EA model (Sec.~\ref{sec:NL}). The equivalence of the two descriptions also requires transforming the observables together with the bonds. The uniform magnetization of the correlated model corresponds, in the original EA model, to the overlap $N^{-1}\sum_iS'_i\sigma^{(0)}_i$ between the spin configuration $S'$ and the reference configuration $\sigma^{(0)}$ (Sec.~\ref{sec:overlap}), not to the uniform magnetization of $S'$.

\subsection{The magnetization is the overlap of two replicas}\label{sec:overlap}

The magnetization of the correlated model,
\begin{equation}
m=\frac1N\sum_iS_i=\frac1N\sum_iS'_i\sigma^{(0)}_i ,
\label{eq:m}
\end{equation}
is, viewed in the gauge of the original EA model, the \textit{overlap} between $S'$ and the reference configuration $\sigma^{(0)}$. This equality holds for each configuration. The three spin variables and their inverse temperatures are related as follows.
\begin{romanlist}
\item $S$: configuration in the gauge of the correlated model. It follows the Boltzmann distribution $e^{\beta\sum\tau_{ij}S_iS_j}/Z_\tau(\beta)$ at inverse temperature $\beta$.
\item $S'_i=S_i\sigma^{(0)}_i$: the same configuration written in the gauge of the original EA model. It follows the Boltzmann distribution $e^{\beta\sum J_{ij}S'_iS'_j}/Z_J(\beta)$ at inverse temperature $\beta$. Since this distribution does not contain $\sigma^{(0)}$, $S'$ is independent of $\sigma^{(0)}$ for fixed $J$.
\item $\sigma^{(0)}$: the reference configuration in the gauge of the original EA model. It is drawn from the Boltzmann distribution $e^{\beta_p\sum J_{ij}\sigma_i\sigma_j}/Z_J(\beta_p)$ at inverse temperature $\beta_p$. In the gauge of the correlated model, it corresponds to the all-up state.
\end{romanlist}
The magnetization $m$ is therefore the overlap between two independent replicas of the EA model with the same bonds $J$, the configuration $S'$ at inverse temperature $\beta$ and the configuration $\sigma^{(0)}$ at inverse temperature $\beta_p$. Writing $q=N^{-1}\sum_iS'_i\sigma_i$ for the overlap between two replicas $S'$ and $\sigma$ of the EA model with the same bonds $J$, we have, in terms of the disorder-averaged distributions,
\begin{align}
P_m(x;\beta,\beta_p)&\equiv\bigl[\langle\delta(x-m)\rangle_{\tau,\beta}\bigr]_{P}
\nonumber\\
&=\EJ\bigl[\langle\delta(x-q)\rangle_{J;\beta,\beta_p}\bigr]\equiv P_q(x;\beta,\beta_p) .
\label{eq:Pm=Pq}
\end{align}
This corresponds to Eq.~(13), the main identity, of Ref.~\cite{Nishimori2024}, where $P_m$ and $P_q$ are denoted by $P_1$ and $P_2$, respectively. On the left-hand side, $[f]_P=\sum_\tau P(\tau)f(\tau)$ is the average over the correlated distribution~\eqref{eq:Ptau}, and $\langle\cdots\rangle_{\tau,\beta}$ is the thermal average at inverse temperature $\beta$ for the bonds $\tau$. On the right-hand side, $\EJ[f]=2^{-N_B}\sum_Jf(J)$ is the average over the bonds of the symmetric $\pm J$ model, and $\langle\cdots\rangle_{J;\beta,\beta_p}$ is the average over two configurations $S'$ and $\sigma$ that independently follow the Boltzmann distributions at inverse temperatures $\beta$ and $\beta_p$ for the bonds $J$,
\begin{equation}
\langle g(S',\sigma)\rangle_{J;\beta,\beta_p}=\sum_{S',\sigma}g(S',\sigma)\,\frac{e^{\beta\sum J_{ij}S'_iS'_j}}{Z_J(\beta)}\,\frac{e^{\beta_p\sum J_{ij}\sigma_i\sigma_j}}{Z_J(\beta_p)} .
\label{eq:two-replica-average}
\end{equation}

We write out the derivation of Eq.~\eqref{eq:Pm=Pq} so that it is clear where the factor $e^{\beta_p\sum\tau_{ij}}/Z_\tau(\beta_p)$ of the correlated distribution goes. Substituting Eq.~\eqref{eq:Ptau}, the left-hand side is
\begin{align}
P_m(x;\beta,\beta_p)={}&\sum_\tau\frac{2^{N-N_B}\,e^{\beta_p\sum\tau_{ij}}}{Z_\tau(\beta_p)}
\nonumber\\
&\times\sum_S\delta\Bigl(x-\frac1N\sum_iS_i\Bigr)\frac{e^{\beta\sum\tau_{ij}S_iS_j}}{Z_\tau(\beta)} .
\end{align}
For fixed $\sigma$, the map $J\mapsto\tau_{ij}=J_{ij}\sigma_i\sigma_j$ is a one-to-one correspondence of bond configurations, so that $\sum_\tau F(\tau)=2^{-N}\sum_\sigma\sum_JF(J\sigma\sigma)$ for any function $F$, where $J\sigma\sigma$ denotes the bonds $J_{ij}\sigma_i\sigma_j$. The factor of the correlated distribution then becomes
\begin{equation}
\frac{e^{\beta_p\sum\tau_{ij}}}{Z_\tau(\beta_p)}=\frac{e^{\beta_p\sum J_{ij}\sigma_i\sigma_j}}{Z_J(\beta_p)} ,
\end{equation}
which is the probability that the configuration $\sigma$ is chosen from the Boltzmann distribution at inverse temperature $\beta_p$ in the system with bonds $J$, i.e., the distribution of the reference configuration. Furthermore, with $S'_i=S_i\sigma_i$, we have $e^{\beta\sum\tau_{ij}S_iS_j}/Z_\tau(\beta)=e^{\beta\sum J_{ij}S'_iS'_j}/Z_J(\beta)$ and $N^{-1}\sum_iS_i=N^{-1}\sum_iS'_i\sigma_i=q$, and hence
\begin{align}
P_m(x;\beta,\beta_p)={}&2^{-N_B}\sum_J\sum_{S',\sigma}\delta(x-q)
\nonumber\\
&\times\frac{e^{\beta\sum J_{ij}S'_iS'_j}}{Z_J(\beta)}\,\frac{e^{\beta_p\sum J_{ij}\sigma_i\sigma_j}}{Z_J(\beta_p)} .
\end{align}
This is precisely the right-hand side $\EJ\bigl[\langle\delta(x-q)\rangle_{J;\beta,\beta_p}\bigr]$ of Eq.~\eqref{eq:Pm=Pq}. In other words, the factor $e^{\beta_p\sum\tau_{ij}}/Z_\tau(\beta_p)$, which on the left-hand side was part of the disorder weight that determines the bonds $\tau$, becomes in the gauge of the EA model the Boltzmann weight of the second replica $\sigma$ and enters the average~\eqref{eq:two-replica-average}. We note that Eq.~\eqref{eq:Pm=Pq} is an equality of distributions after the average over the disorder; it does not state that the distribution of the magnetization for a single fixed $\tau$ is equal to the distribution of the overlap in the EA model.

For $\gamma\neq0$, the generation procedure of Sec.~\ref{sec:gamma} gives in the same way Eq.~\eqref{eq:Pm=Pq} with $\EJ$ replaced by the average over $P_{\rm EA}(J;\gamma)$: the magnetization is the overlap of two replicas of EA($\gamma$) [Eq.~(14) of NOO].

\subsection{On and off the Nishimori line}\label{sec:NL}

In what follows, we say that the order vanishes if $P_m$ concentrates on $\delta(x)$ in the thermodynamic limit. Since $|m|\leq1$, this is equivalent to $[\langle m^2\rangle]_P\to0$. In a symmetric finite system at zero field, $[\langle m\rangle]_P=0$ by the symmetry $S\to-S$, and the second moment or the absolute value is used to detect the order. Equation~\eqref{eq:Pm=Pq} gives $[\langle m^2\rangle]_P=\EJ\langle q^2\rangle$ and $[\langle|m|\rangle]_P=\EJ\langle|q|\rangle$.

\paragraph{On the Nishimori line.}
The line $\beta=\beta_p$ is the Nishimori line of the correlated model~\cite{Nishimori2024}. On this line, the two replicas are at the same temperature, and the distribution of the magnetization is equal to the distribution of the overlap of the EA model at the same temperature. In this sense, the ferromagnetic phase of the correlated model on the Nishimori line is the spin-glass phase of the EA model in disguise: since $[\langle m^2\rangle]_P=\EJ\langle q^2\rangle$ at the same temperature, the ferromagnetic order of the correlated model exists if and only if the spin-glass order of the EA model does. If the overlap distribution of the EA model has a nontrivial form of the replica-symmetry-breaking (RSB) type~\cite{Parisi1979,Mezard1987}, this form is transferred to the distribution of the magnetization of the correlated model~\cite{Nishimori2024}. To relate the average of the magnetization to the spin-glass order parameter of the EA model, one has to specify the boundary conditions or the choice of states so that it does not vanish trivially by symmetry (fixed boundary conditions are used in Refs.~\cite{Nishimori2024,Nishimori2025}).

\paragraph{Off the Nishimori line.}
For $\beta\neq\beta_p$, the distribution of the magnetization is equal to the distribution of the overlap between two replicas at different temperatures. The concentration of the magnetization distribution of the correlated model on $\delta(x)$ is therefore equivalent to the concentration on $\delta(x)$ of the distribution of the overlap between the temperatures $\beta$ and $\beta_p$. When both temperatures are in the spin-glass phase at zero field, the latter is nothing but temperature chaos~\cite{Bray1987,Kondor1989}. (If one of the two temperatures is in the paramagnetic phase, the overlap vanishes without any assumption, since $(\EJ\langle q^2\rangle_{J;\beta,\beta_p})^2\leq\EJ\langle q^2\rangle_{J;\beta,\beta}\,\EJ\langle q^2\rangle_{J;\beta_p,\beta_p}$ by the Cauchy--Schwarz inequality; this is unrelated to temperature chaos.) If temperature chaos is assumed for all pairs of distinct temperatures in the spin-glass phase, i.e., $\lim_{N\to\infty}\EJ\langle q^2\rangle_{J;\beta,\beta_p}=0$ for all $\beta\neq\beta_p$ in this phase, the magnetization of the correlated model vanishes away from the Nishimori line, and the ferromagnetic order is confined to the Nishimori line~\cite{Nishimori2024}, a line rather than a region of the phase diagram. If chaos is assumed only for a particular pair of temperatures, the conclusion holds only at that point.

\section{Destruction of the ferromagnetic phase by random fields: conditions and intuition}\label{sec:field}

\subsection{Magnetization and the overlap between zero-field and in-field configurations}\label{sec:overlap-field}

In this section and in Sec.~\ref{sec:IsingXY}, we consider the case $\gamma=0$, i.e., the distribution~\eqref{eq:Ptau} and, in Sec.~\ref{sec:IsingXY}, its $XY$ analog. We add a random field to the correlated model, so that the Boltzmann weight becomes $e^{\beta(\sum\tau_{ij}S_iS_j+h\sum_i\mu_iS_i)}$. Here, $h$ is the strength of the field (Ref.~\cite{Nishimori2025} uses the dimensionless field $\beta h$), and the directions $\mu_i=\pm1$ are random variables that take the two values with probability $1/2$ each, independently from site to site and independently of the bonds and of the reference configuration. In the gauge of the original EA model, the field term becomes $h\sum_i\eta_iS'_i$ with $\eta_i=\mu_i\sigma^{(0)}_i$. Since the $\mu_i$ are symmetric, mutually independent, and independent of $(J,\sigma^{(0)})$, the $\eta_i$ have the same properties. For $\gamma=0$, where the distribution of $J$ is symmetric, a further gauge transformation $S'_i\to\eta_iS'_i$, $\sigma^{(0)}_i\to\eta_i\sigma^{(0)}_i$, $J_{ij}\to J_{ij}\eta_i\eta_j$ makes the field uniform, keeps the bonds independent and symmetrically distributed, and leaves the overlap unchanged. (For $\gamma\neq0$, the rewriting in the gauge of the EA model still holds, so that the magnetization is the overlap between a zero-field configuration and a configuration in the random field $h\eta_i$, for bonds drawn from $P_{\rm EA}(J;\gamma)$; the field cannot then be made uniform without changing the distribution of the bonds.) Hence, on the Nishimori line $\beta=\beta_p$,
\begin{equation}
P_m(x;\beta,h;\beta,0)=P_q(x;\beta,h;\beta,0) ,
\end{equation}
where the arguments are the inverse temperature and field of the observed configuration, followed by those of the reference configuration. The left-hand side is the distribution of the magnetization of the correlated model under the random field, averaged over the bonds and over the directions $\mu_i$ of the field, and the right-hand side is the distribution of the overlap, in the same EA model, between a configuration in the uniform field $h$ and a configuration at zero field~\cite{Nishimori2025}. This is an identity of disorder-averaged distributions and holds in any dimension.

Consequently, if the right-hand side concentrates on $\delta(x)$ in the thermodynamic limit for any fixed $h>0$, that is, if the overlap between zero-field and in-field configurations vanishes, then the ferromagnetic phase of the correlated model is destroyed by the random field (the limit $h\to0$ is taken after the thermodynamic limit). Physically, the vanishing of this overlap means that the spin-glass phase at zero field either exhibits magnetic-field chaos~\cite{Kondor1989,Ritort1994,Billoire2003,AguilarJanita2024} or is destroyed by the field. However, ``absence of a spin-glass phase in a field'' and ``vanishing of the overlap with zero-field configurations'' are different propositions and cannot be regarded as equivalent in general. Below, we consider in the droplet picture~\cite{Fisher1986,Fisher1988} why this overlap is expected to vanish.

\subsection{Cost of a domain wall}\label{sec:wall}

In this and the next subsection, we develop an argument of the Imry--Ma type~\cite{Imry1975}: the cost of reversing the spins in a region of linear size $\ell$ (this subsection) is compared with the energy gained from the random field by this reversal (Sec.~\ref{sec:gain}). The energy difference for reversing the spins of a configuration $S$ in a region $D$ of linear size $\ell$ is $2\sum_{ij\in\partial D}\tau_{ij}S_iS_j$, summed over the bonds crossing the boundary (starting from the all-up state, $2\sum_{ij\in\partial D}\tau_{ij}$). To discuss the stability of a phase at finite temperature, one has to consider the free energy of the optimal excitation, including thermal fluctuations. We call this free energy, which does not include the field term, the cost of the reversal and denote it by $\Delta F(\ell)$.
\begin{romanlist}
\item \textit{Pure ferromagnet.} All bonds are $\tau_{ij}=+1$, so every bond of the wall costs energy wherever the wall is placed, and at sufficiently low temperatures the cost is proportional to the area, $\Delta F(\ell)\propto\ell^{d-1}$ (positive interface tension). In the ferromagnetic phase of an EA model with an independent distribution biased toward ferromagnetic bonds, a positive interface tension is also believed to remain.
\item \textit{Correlated model.} The order of the correlated model is the spin-glass order viewed in a different gauge, and its excitations are spin-glass droplets. In the droplet picture, the cost is believed to grow only as $\Delta F(\ell)\sim\ell^{\theta}$, because the contributions of satisfied and unsatisfied bonds cancel, and Fisher and Huse estimated $\theta\leq(d-1)/2$~\cite{Fisher1988}. The exponent $\theta$ is not determined merely by the existence of unsatisfied bonds; this is a picture of the structure of the optimal excitations.
\end{romanlist}

\subsection{Comparison with the gain from random fields}\label{sec:gain}

Since the field is independent of the reference configuration, the energy gained by reversing the region $D$ in the direction favored by the field is of the order of the square root of the volume, i.e., of order $h\ell^{d/2}$. This estimate assumes that the candidate droplets for reversal are determined independently of the field and that a finite fraction of the spins in them change direction (the factor due to the spin-glass order parameter is omitted). We call this energy the gain. The field wins, i.e., regions are reversed in the direction favored by the field, if the gain exceeds the cost on large scales,
\begin{equation}
\Delta F(\ell)\ll h\ell^{d/2}\qquad(\ell\to\infty).
\label{eq:IM}
\end{equation}
\begin{romanlist}
\item \textit{Pure ferromagnet.} At sufficiently low temperatures and for a weak random field, the comparison of $\ell^{d-1}$ with $h\ell^{d/2}$ suggests that the surface energy wins for $d\geq3$ (the Imry--Ma argument~\cite{Imry1975}); the survival of ferromagnetic order in a weak random field for $d\geq3$ has been established rigorously~\cite{Imbrie1985,BricmontKupiainen1988}.
\item \textit{Correlated model.} In the droplet picture, the comparison is between $\ell^{\theta}$ and $h\ell^{d/2}$. Since $\theta<d/2$, Eq.~\eqref{eq:IM} holds in any dimension: the field wins on scales larger than $\ell_h\sim h^{-1/(d/2-\theta)}$, the droplets favored by the field are reversed, and the overlap with the reference configuration, i.e., the magnetization $m$, is expected to vanish.
\end{romanlist}
This is the argument of Fisher and Huse that a uniform field destroys the spin-glass phase of the EA model in the droplet picture~\cite{Fisher1988}, reread in the gauge of the correlated model. In the RSB picture, the same vanishing of the magnetization follows if the magnetic-field chaos expected there exists; RSB alone, however, does not imply magnetic-field chaos. In either case, this is a physical explanation based on pictures of droplet scaling or magnetic-field chaos, and not a direct proof, in all dimensions, that the distribution of the magnetization concentrates on $\delta(x)$. The form of the Imry--Ma energy comparison thus remains the same, with the cost of a wall replaced from the value for a ferromagnet by the value for a spin glass. We note that this comparison in terms of $\ell$ concerns short-range lattice models and does not apply directly to long-range or mean-field models. The cases in which the comparison can be turned into a proof are discussed in Sec.~\ref{sec:IsingXY}.

\begin{table}[b]
\caption{\label{tab:costs}Comparison of the cost of reversing a region of linear size $\ell$ with the gain $h\ell^{d/2}$ from random fields, Eq.~\eqref{eq:IM}. The column ``Cost'' gives the actual cost $\Delta F(\ell)$ of the optimal excitation in the first two rows (for the correlated model, an estimate within the droplet picture) and the upper bound $U(\ell)$ on the change of the interaction energy, Eq.~\eqref{eq:bound}, in the last two. The bound leads to a proof where it is smaller than the gain and to no conclusion elsewhere. Marginal dimensions are omitted.}
\begin{ruledtabular}
\begin{tabular}{llll}
Model & Cost & Kind of cost & Field wins for\\
\hline
Pure Ising ferromagnet & $\ell^{d-1}$ & actual & $d<2$\\
Correlated Ising model & $\ell^{\theta}$ & actual (droplet) & any $d$\\
Correlated Ising model & $\ell^{d-1}$ & upper bound & $d<2$\\
Correlated $XY$ model & $\ell^{d-2}$ & upper bound & $d<4$\\
\end{tabular}
\end{ruledtabular}
\end{table}
\section{Ising versus \texorpdfstring{$XY$}{XY}: the range covered by proofs}\label{sec:IsingXY}

Table~\ref{tab:costs} lists the forms of the cost used in the comparison~\eqref{eq:IM}. In the first two rows, it is the actual cost $\Delta F(\ell)$ of the optimal excitation of Secs.~\ref{sec:wall} and \ref{sec:gain}: $\ell^{d-1}$ for the pure ferromagnet and $\ell^{\theta}$ for the correlated model. The latter is an estimate within the droplet picture, and the conclusion that the field wins in any dimension is therefore not a proof. In the last two rows, the actual cost is replaced by an upper bound $U(\ell)$ on the change of the interaction energy, i.e., the energy without the field term, caused by the reversal, valid for every spin configuration and every bond configuration. Since the cost cannot exceed such a bound,
\begin{equation}
\Delta F(\ell)\leq U(\ell).
\label{eq:bound}
\end{equation}
If $U(\ell)\ll h\ell^{d/2}$, Eq.~\eqref{eq:IM} follows from Eq.~\eqref{eq:bound}, without any assumption on the nature of the spin-glass phase or on magnetic-field chaos. If $U(\ell)$ exceeds $h\ell^{d/2}$, Eq.~\eqref{eq:bound} gives no conclusion; in particular, it does not imply that the order survives. The bound $U(\ell)$ is much larger than the cost $\ell^{\theta}$ expected in the droplet picture, so that the field is shown to win only in low dimensions (last column of Table~\ref{tab:costs}). We now explain the bounds for Ising and $XY$ spins.
\begin{figure}[t]
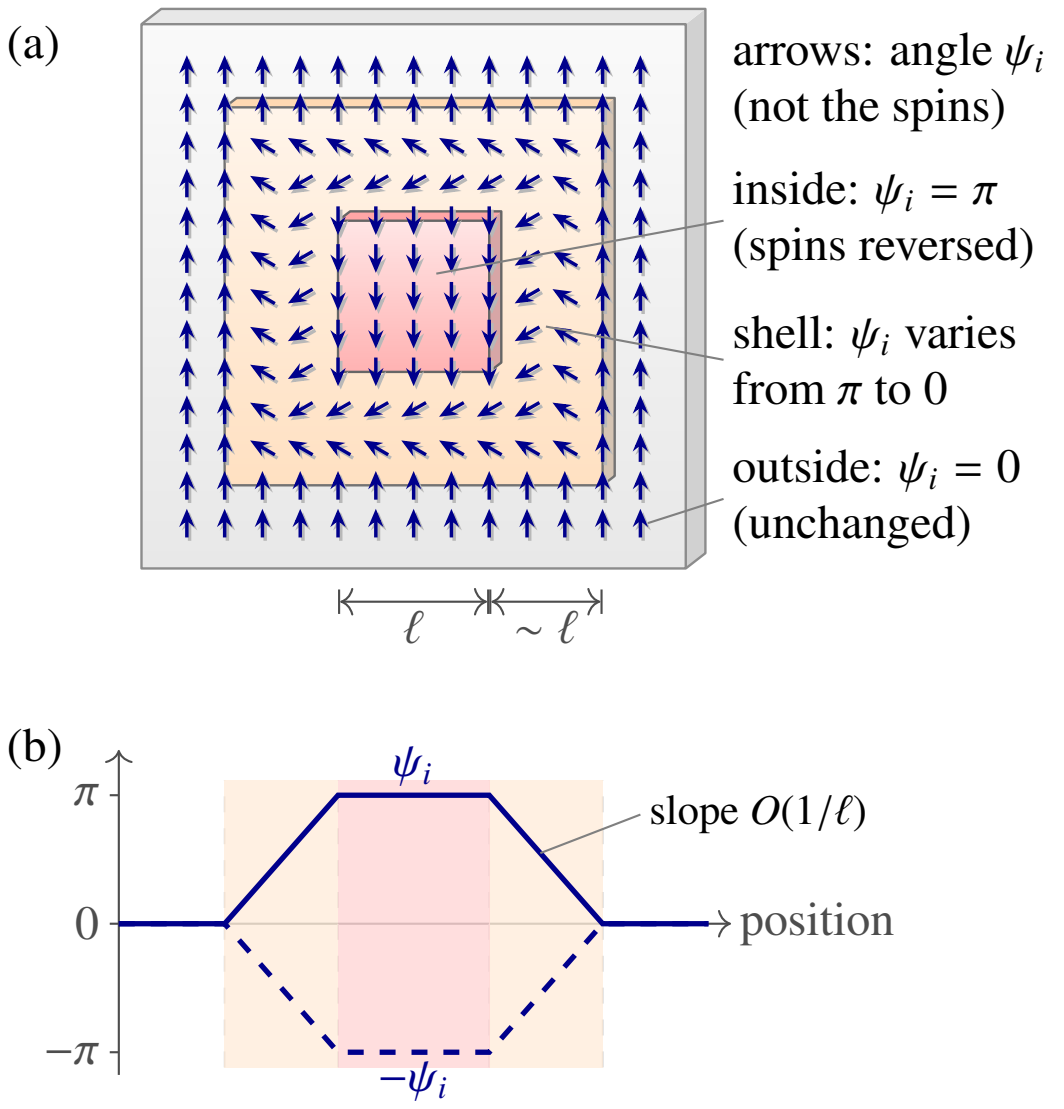

\centering

\caption{\label{fig:twist}Twist used in Sec.~\ref{sec:IsingXY} for the $XY$ gauge glass. The figure shows the rotation angle $\psi_i$ of the twist, not the spins: the same twist $\theta_i\to\theta_i+\psi_i$ is applied to every spin configuration, however irregular, and for every realization of the phase shifts $B_{ij}$. (a) The spins are rotated by $\psi_i=\pi$, i.e., reversed, inside a region of linear size $\ell$ (inner square) and unchanged ($\psi_i=0$) outside a larger region of linear size of order $\ell$ (outer square); in the shell between them, $\psi_i$ varies smoothly from $\pi$ to $0$. The arrow at each site is the upward direction rotated by $\psi_i$. (b) Profile of $\psi_i$ (solid line) along a horizontal line through the center, together with that of the opposite twist $-\psi_i$ (dashed line). The difference $\Delta_{ij}=\psi_i-\psi_j$ across a bond is of order $1/\ell$ in the shell and zero elsewhere, and has opposite signs for the two twists. Since rotations by $\pi$ and $-\pi$ are the same, the two twists give the same configuration inside the inner region.}
\end{figure}
\begin{romanlist}
\item \textit{Ising.} Since the energy change $2\sum_{ij\in\partial D}\tau_{ij}S_iS_j$ of Sec.~\ref{sec:wall} cannot exceed twice the number of boundary bonds, the available bound is $U(\ell)\sim\ell^{d-1}$, the same as the cost for the pure ferromagnet. It is smaller than the gain $h\ell^{d/2}$ only for $d<2$; the marginal case $d=2$ requires further arguments such as those of Aizenman and Wehr~\cite{AizenmanWehr1989,AizenmanWehr1990}. Since the two-dimensional EA model is believed to have no spin-glass phase at finite temperature~\cite{Hartmann2001,Jorg2006}, this gives nothing new. For $d\geq3$, the bound exceeds the gain, and the destruction of the ferromagnetic order expected in Sec.~\ref{sec:gain} remains the conditional statement of Sec.~\ref{sec:overlap-field}.
\item \textit{$XY$.} For continuous spins, a smaller bound is available. The correlated model is here the $XY$ gauge glass in a random field, with the Hamiltonian~\cite{Okuyama2026}
\begin{equation}
H(h)=-\sum_{\langle ij\rangle}\cos(\theta_i-\theta_j-B_{ij})-h\sum_ir_i\cos(\theta_i-\phi_i),
\label{eq:XYgg}
\end{equation}
where $\theta_i$ is the angle of the spin at site $i$, and the Boltzmann weight is $e^{-\beta H(h)}$. The phase shifts $B_{ij}$ follow the distribution proportional to $e^{\beta_p\sum_{\langle ij\rangle}\cos B_{ij}}/Z_0(B;\beta_p)$, the $XY$ analog of Eq.~\eqref{eq:Ptau}, where $Z_0$ is the partition function at zero field; as in Sec.~\ref{sec:overlap-field}, we consider the Nishimori line $\beta_p=\beta$. The random field is an isotropic Gaussian field~\cite{Okuyama2026}: its components $r_i\cos\phi_i$ and $r_i\sin\phi_i$ are Gaussian variables with zero mean and unit variance, independent of each other, from site to site, and of the bonds. Equivalently, the directions $\phi_i$ are uniformly distributed in $[0,2\pi)$, and the amplitudes $r_i\geq0$ follow the Rayleigh distribution $p(r)=re^{-r^2/2}$. We write the energy of the bond $ij$ in the first term as $e(x_{ij})=-\cos x_{ij}$ with $x_{ij}=\theta_i-\theta_j-B_{ij}$.

Consider a slow twist that reverses the spins, i.e., rotates them by $\pi$, inside a region of linear size $\ell$ (the inner region) and leaves them unchanged outside a larger region of linear size of order $\ell$ (the outer region). The value $\pi$ is essential for the cancellation below. The twist changes the angles as $\theta_i\to\theta_i+\psi_i$, with $\psi_i=\pi$ in the inner region, $\psi_i=0$ outside the outer region, and $\psi_i$ varying smoothly in the shell between them (Fig.~\ref{fig:twist}). The variable of each bond then changes as $x_{ij}\to x_{ij}+\Delta_{ij}$ with $\Delta_{ij}=\psi_i-\psi_j$, which is of order $1/\ell$ on the $O(\ell^d)$ bonds of the shell and zero elsewhere.

The energy change of a bond,
\[
e(x_{ij}+\Delta_{ij})-e(x_{ij})=\Delta_{ij}\sin x_{ij}+O(\Delta_{ij}^2),
\]
is of first order in $\Delta_{ij}$, with a sign that depends on the spin configuration and the phase shift through $\sin x_{ij}$. In the gauge glass, the angles $\theta_i$ are irregular because of the phase shifts, and this sign varies from bond to bond. Bounding $|\Delta_{ij}\sin x_{ij}|$ by $|\Delta_{ij}|$ bond by bond gives only a bound of order $\ell^d\times\ell^{-1}=\ell^{d-1}$, as for Ising spins.

The first-order terms are canceled by applying to the same spin configuration the twist $\psi_i$ and the opposite twist $-\psi_i$ [Fig.~\ref{fig:twist}(b)], as in the Mermin--Wagner argument~\cite{MerminWagner1966}. The variable $x_{ij}$ is determined by the configuration before the twist and is common to the two twists, whereas $\Delta_{ij}$ has opposite signs. The first-order terms $+\Delta_{ij}\sin x_{ij}$ and $-\Delta_{ij}\sin x_{ij}$ therefore cancel on each bond, whatever the value of $x_{ij}$:
\begin{equation}
e(x+\Delta)+e(x-\Delta)-2e(x)=2\cos x\,(1-\cos\Delta)\leq\Delta^2
\label{eq:twist}
\end{equation}
for any $x$ and $\Delta$. This cancellation is exact for each spin configuration and each realization of the phase shifts $B_{ij}$, however irregular; it does not rely on an average over $B_{ij}$ or over the spin configurations. Summed over the $O(\ell^d)$ bonds of the shell with $\Delta_{ij}=O(1/\ell)$, the sum of the energy changes of the two twists is bounded by $O(\ell^{d-2})$, uniformly in the spin configuration and in the bonds, which gives $U(\ell)\sim\ell^{d-2}$.

If one of the two twists raises the energy at first order, the other lowers it by the same amount, so that, for every spin configuration, at least one of them costs no more than half of this bound. Since rotations by $+\pi$ and $-\pi$ are the same, the two twisted configurations coincide in the inner region, where the spins are reversed, and differ only in the shell: the region can thus always be reversed at a cost of at most $O(\ell^{d-2})$. At finite temperature, the corresponding statement for the free energy follows from the inequality $e^{-\beta E_+}+e^{-\beta E_-}\geq2e^{-\beta(E_++E_-)/2}$ for the energies $E_\pm$ of the two twisted configurations: since each twist is a one-to-one map of the spin configurations, integrating this inequality over the configurations gives a comparison of partition functions. The bound $U(\ell)\sim\ell^{d-2}$ is smaller than the gain $h\ell^{d/2}$ for $d<4$, and $d=4$ is the marginal case.

The bound $U(\ell)$ concerns the interaction energy. The field term of Eq.~\eqref{eq:XYgg} also changes in the shell, and for a fixed realization of the field it is not bounded in the same way. In the proof of Ref.~\cite{Okuyama2026}, the random field is rotated together with the spins, which leaves the field term unchanged, and the rotated field has the same distribution as the original one because the distribution of the field is isotropic. This step requires only the isotropy; the remaining steps of the proof use the Gaussian distribution of the field.

Theorem~1 of Ref.~\cite{Okuyama2026}, obtained by extending the method of Dario, Harel, and Peled~\cite{Dario2024} to bond-dependent interactions, makes this argument rigorous: for $d\leq4$, at any finite temperature and any fixed $h>0$, the second moment of the magnetization of the correlated model~\eqref{eq:XYgg}, averaged over the random field, vanishes in the thermodynamic limit, uniformly in the bonds.

Through the gauge identity of Ref.~\cite{Okuyama2026}, the second moment of the overlap between equilibrium configurations of the $XY$ gauge glass at zero field and in a field then vanishes, for a field in a fixed direction with amplitudes following the Rayleigh distribution. It follows that, for $d\leq4$, such a field either induces magnetic-field chaos in the spin-glass phase or destroys spin-glass order. In other words, Ref.~\cite{Okuyama2026} proves that, for $d\leq4$, if a spin-glass phase persists in a field of this type, it necessarily exhibits magnetic-field chaos. A field of strictly uniform amplitude is outside this conclusion.
\end{romanlist}
For $d>4$, the bound by twisting exceeds the gain and gives no conclusion. What is needed there is the actual cost $\Delta F(\ell)$, as in Secs.~\ref{sec:wall} and \ref{sec:gain}, which is where the droplet and RSB pictures differ. Numerical tests of the ferromagnetic order of the correlated model in $d\geq5$, in which the bonds can be generated by the $XY$ analog of the procedure of Sec.~\ref{sec:procedure}, have been proposed in Ref.~\cite{Okuyama2026}.

\section{Discussion}\label{sec:discussion}

The correspondences of Secs.~\ref{sec:identity} and \ref{sec:field} form a dictionary that translates questions about spin-glass phases (RSB, temperature chaos, and magnetic-field chaos) into questions about the ferromagnetic order of the correlated model. A dictionary is powerful when something can be said about one side by other means, as in the following three cases.
\begin{romanlist}
\item \textit{Logical relation between temperature chaos and reentrant transitions}~\cite{Nishimori2025a}. Varying the frustration through $\gamma$ independently of $\beta_p$ and combining the dictionary with the structure of the resulting 3D phase diagram with axes $(\gamma,\beta_p,1/\beta)$, NOO showed that the boundary between the ferromagnetic and spin-glass phases of the EA model is not reentrant if a spin-glass phase exists at finite temperature without temperature chaos. The contrapositive states that, if a spin-glass phase exists at finite temperature and the boundary is reentrant, there is at least one pair of distinct temperatures in the spin-glass phase for which the overlap vanishes~\cite{Nishimori2026pspin}. This is a necessary condition for temperature chaos, weaker than the vanishing of the overlap for all pairs of temperatures assumed in Sec.~\ref{sec:NL}. For the fully connected $p$-spin model, a reentrant transition near the triple point has been shown by an analytical expansion within the replica method for the values of $p$ examined, and the necessary condition for temperature chaos then follows from this relation~\cite{Nishimori2026pspin}. A similar analysis has been carried out for the Potts gauge glass (in preparation).
\item \textit{Numerical study of the model of Kitatani}~\cite{Oshima2026}. With the correlated disorder generated by the procedure of Sec.~\ref{sec:gamma}, implemented with tensor-network proposals and importance reweighting, OAH studied numerically, for systems of up to $256\times256$ spins, a possible intermediate Mattis-like spin-glass phase, with spin-glass order but without ferromagnetic order, related to the reentrant transition of the two-dimensional EA model. Their finite-size scaling of the Binder ratios places the spin-glass and ferromagnetic transitions at slightly different temperatures, but the system-size dependence of the squared order parameters shows no clear range with spin-glass order and without ferromagnetic order. The existence of an intermediate phase is thus not established, although a very narrow one cannot be ruled out.
\item \textit{The $XY$ gauge glass}~\cite{Okuyama2026}. Here the side of ferromagnetic order is accessible to rigorous analysis. The destruction of the ferromagnetic order of the correlated model by a random field is proved for $d\leq4$ (Sec.~\ref{sec:IsingXY}), and the dictionary turns this into the statement that spin-glass order in a field of the type considered there requires magnetic-field chaos. The proof can be applied because the random field is independent of the bonds after the gauge transformation and the estimates are uniform in the bonds, so that they can be averaged over the correlated distribution.
\end{romanlist}

Refs.~\cite{Nishimori2024,Nishimori2025,Nishimori2025a} derived their results from sums over $\sigma$ in gauge-transformed expressions. The reading of the model as the EA model with an embedded equilibrium configuration is contained in these sums but was not stated explicitly before OAH. Once stated, the correspondence may seem almost obvious. The gauge identities behind it have nevertheless served as the foundation for the highly nontrivial results of items (i) and (iii), which, to our knowledge, have not been obtained by other approaches.

\section{Summary}\label{sec:summary}

The correlated disorder is obtained by drawing one reference configuration from the equilibrium distribution of the symmetric $\pm J$ model and choosing the gauge in which it looks like the all-up state; this is the case $\gamma=0$ of the procedure given by OAH for the generalized distribution of NOO. The correlated model is in this sense a spin-glass counterpart of the Mattis model, in which a ferromagnet is written in a random gauge unrelated to the bonds. In the generalization of NOO, $\gamma$ determines the probabilities of the gauge equivalence classes, and hence the frustration statistics, whereas $\beta_p$ determines the conditional distribution within each class, i.e., the choice of gauge. The magnetization of the correlated model is the overlap with the reference configuration. These statements hold exactly in finite systems.

The response of the ferromagnetic order of the correlated model to temperature and field is therefore determined by the overlap between two replicas of the EA model under different conditions. If the EA model exhibits temperature chaos for all pairs of distinct temperatures in the spin-glass phase, the ferromagnetic order is confined to the Nishimori line, and if it exhibits magnetic-field chaos, the ferromagnetic order is destroyed by random fields in any dimension; the droplet picture gives an intuitive reason for the latter. The fragility of the ferromagnetic order is thus not a general effect of correlations in the disorder but a consequence of this particular embedding.

This fragility is, at the same time, what makes the correspondence fruitful: properties of ferromagnetic order that can be established by other means translate into properties of spin-glass phases. For the $XY$ gauge glass, the destruction of the ferromagnetic order of the correlated model by random fields, proved for $d\leq4$, yields the nontrivial rigorous result that, in the original model with independent bonds, a field in a fixed direction with site-dependent amplitudes either induces magnetic-field chaos in the spin-glass phase or destroys spin-glass order~\cite{Okuyama2026}. By establishing this correspondence between spin-glass phases and the ferromagnetic order of models with correlated disorder, the series of studies discussed here~\cite{Nishimori2024,Nishimori2025,Nishimori2025a,Oshima2026,Okuyama2026} has opened a new approach to the analysis of spin glasses, which we expect to lead to further results on other properties of spin-glass phases and on other models with gauge symmetry.

\protectedsectionskip
\section*{Methods: Use of generative-AI tools}\label{sec:methods}

The physical interpretation presented in this paper was developed through discussions between the author and the generative-AI tools Claude Opus 5.5 and Claude Fable 5.2 (Anthropic). The author posed the questions to be addressed, in particular the meaning of the correlated distributions~\eqref{eq:Ptau} and \eqref{eq:Pgamma} and their relation to the generation procedure of OAH and to the Mattis model, and directed the discussion. Claude Opus 5.5 and Claude Fable 5.2 were used to propose explanations and to work out derivations, such as those of Secs.~\ref{sec:construction}--\ref{sec:identity}, in response to the author's questions and instructions. Figures~\ref{fig:procedure}--\ref{fig:twist} were drawn with TikZ code prepared with the help of the Claude models under the author's direction and checked by the author; in Fig.~\ref{fig:procedure}, the reference configuration was drawn from the Boltzmann distribution computed exactly for a $4\times4$ lattice. No generative image model was used.

The author examined and independently verified each step of the AI-generated derivations. GPT-6.1 Sol (OpenAI) was additionally used to independently examine the derivations and identify possible errors or inconsistencies. Issues found either by the author or through this cross-checking process were fed back to the Claude models for correction, after which the revised calculations were again examined by the author. This iterative procedure was repeated throughout the analytical development.

All decisions concerning the physical approach, the acceptance or rejection of AI-generated results, and their interpretation were made by the author, who takes full responsibility for the scientific content of this work.

\begin{acknowledgments}
Claude Opus 5.5 and Claude Fable 5.2 (Anthropic), with additional use of GPT-6.1 Sol (OpenAI), were used extensively in writing the manuscript through iterative interaction with the author (their use in developing the content is described in the Methods section). The author gave detailed and repeated instructions for revision, correction, and restructuring, and reviewed and edited the resulting text. The author takes full responsibility for the final wording and presentation of the manuscript.
\end{acknowledgments}

\protectedsectionskip
\section*{Data Availability Statement}
This manuscript has no associated data: this is a theoretical study, and no datasets were generated or analysed. For the small example in Fig.~\ref{fig:procedure}, the bonds $J_{ij}$, the reference configuration $\sigma^{(0)}$, and the resulting bonds $\tau_{ij}$ are all shown in the figure.

\bibliography{references}

\begin{thebibliography}{33}%
\makeatletter
\providecommand \@ifxundefined [1]{%
 \@ifx{#1\undefined}
}%
\providecommand \@ifnum [1]{%
 \ifnum #1\expandafter \@firstoftwo
 \else \expandafter \@secondoftwo
 \fi
}%
\providecommand \@ifx [1]{%
 \ifx #1\expandafter \@firstoftwo
 \else \expandafter \@secondoftwo
 \fi
}%
\providecommand \natexlab [1]{#1}%
\providecommand \enquote  [1]{``#1''}%
\providecommand \bibnamefont  [1]{#1}%
\providecommand \bibfnamefont [1]{#1}%
\providecommand \citenamefont [1]{#1}%
\providecommand \href@noop [0]{\@secondoftwo}%
\providecommand \href [0]{\begingroup \@sanitize@url \@href}%
\providecommand \@href[1]{\@@startlink{#1}\@@href}%
\providecommand \@@href[1]{\endgroup#1\@@endlink}%
\providecommand \@sanitize@url [0]{\catcode `\\12\catcode `\$12\catcode
  `\&12\catcode `\#12\catcode `\^12\catcode `\_12\catcode `\%12\relax}%
\providecommand \@@startlink[1]{}%
\providecommand \@@endlink[0]{}%
\providecommand \url  [0]{\begingroup\@sanitize@url \@url }%
\providecommand \@url [1]{\endgroup\@href {#1}{\urlprefix }}%
\providecommand \urlprefix  [0]{URL }%
\providecommand \Eprint [0]{\href }%
\providecommand \doibase [0]{https://doi.org/}%
\providecommand \selectlanguage [0]{\@gobble}%
\providecommand \bibinfo  [0]{\@secondoftwo}%
\providecommand \bibfield  [0]{\@secondoftwo}%
\providecommand \translation [1]{[#1]}%
\providecommand \BibitemOpen [0]{}%
\providecommand \bibitemStop [0]{}%
\providecommand \bibitemNoStop [0]{.\EOS\space}%
\providecommand \EOS [0]{\spacefactor3000\relax}%
\providecommand \BibitemShut  [1]{\csname bibitem#1\endcsname}%
\let\auto@bib@innerbib\@empty
\bibitem [{\citenamefont {Nishimori}(1981)}]{Nishimori1981}%
  \BibitemOpen
  \bibfield  {author} {\bibinfo {author} {\bibfnamefont {H.}~\bibnamefont
  {Nishimori}},\ }\bibfield  {title} {\bibinfo {title} {Internal energy,
  specific heat and correlation function of the bond-random {Ising} model},\
  }\href {https://doi.org/10.1143/PTP.66.1169} {\bibfield  {journal} {\bibinfo
  {journal} {Prog. Theor. Phys.}\ }\textbf {\bibinfo {volume} {66}},\ \bibinfo
  {pages} {1169} (\bibinfo {year} {1981})}\BibitemShut {NoStop}%
\bibitem [{\citenamefont {Nishimori}(2001)}]{Nishimori2001}%
  \BibitemOpen
  \bibfield  {author} {\bibinfo {author} {\bibfnamefont {H.}~\bibnamefont
  {Nishimori}},\ }\href@noop {} {\emph {\bibinfo {title} {{Statistical Physics
  of Spin Glasses and Information Processing: An Introduction}}}}\ (\bibinfo
  {publisher} {Oxford University Press},\ \bibinfo {address} {Oxford},\
  \bibinfo {year} {2001})\BibitemShut {NoStop}%
\bibitem [{\citenamefont {Nishimori}(2024)}]{Nishimori2024}%
  \BibitemOpen
  \bibfield  {author} {\bibinfo {author} {\bibfnamefont {H.}~\bibnamefont
  {Nishimori}},\ }\bibfield  {title} {\bibinfo {title} {{Anomalous distribution
  of magnetization in an Ising spin glass with correlated disorder}},\ }\href
  {https://doi.org/10.1103/PhysRevE.110.064108} {\bibfield  {journal} {\bibinfo
   {journal} {Phys. Rev. E}\ }\textbf {\bibinfo {volume} {110}},\ \bibinfo
  {pages} {064108} (\bibinfo {year} {2024})}\BibitemShut {NoStop}%
\bibitem [{\citenamefont {Nishimori}(2025)}]{Nishimori2025}%
  \BibitemOpen
  \bibfield  {author} {\bibinfo {author} {\bibfnamefont {H.}~\bibnamefont
  {Nishimori}},\ }\bibfield  {title} {\bibinfo {title} {{Instability of the
  ferromagnetic phase under random fields in an Ising spin glass with
  correlated disorder}},\ }\href {https://doi.org/10.1103/PhysRevE.111.044109}
  {\bibfield  {journal} {\bibinfo  {journal} {Phys. Rev. E}\ }\textbf {\bibinfo
  {volume} {111}},\ \bibinfo {pages} {044109} (\bibinfo {year}
  {2025})}\BibitemShut {NoStop}%
\bibitem [{\citenamefont {Edwards}\ and\ \citenamefont
  {Anderson}(1975)}]{Edwards1975}%
  \BibitemOpen
  \bibfield  {author} {\bibinfo {author} {\bibfnamefont {S.~F.}\ \bibnamefont
  {Edwards}}\ and\ \bibinfo {author} {\bibfnamefont {P.~W.}\ \bibnamefont
  {Anderson}},\ }\bibfield  {title} {\bibinfo {title} {Theory of spin
  glasses},\ }\href {https://doi.org/10.1088/0305-4608/5/5/017} {\bibfield
  {journal} {\bibinfo  {journal} {J. Phys. F}\ }\textbf {\bibinfo {volume}
  {5}},\ \bibinfo {pages} {965} (\bibinfo {year} {1975})}\BibitemShut {NoStop}%
\bibitem [{\citenamefont {Nishimori}\ \emph {et~al.}(2025)\citenamefont
  {Nishimori}, \citenamefont {Ohzeki},\ and\ \citenamefont
  {Okuyama}}]{Nishimori2025a}%
  \BibitemOpen
  \bibfield  {author} {\bibinfo {author} {\bibfnamefont {H.}~\bibnamefont
  {Nishimori}}, \bibinfo {author} {\bibfnamefont {M.}~\bibnamefont {Ohzeki}},\
  and\ \bibinfo {author} {\bibfnamefont {M.}~\bibnamefont {Okuyama}},\
  }\bibfield  {title} {\bibinfo {title} {{Temperature chaos as a logical
  consequence of the reentrant transition in spin glasses}},\ }\href
  {https://doi.org/10.1103/qp1w-qcbs} {\bibfield  {journal} {\bibinfo
  {journal} {Phys. Rev. E}\ }\textbf {\bibinfo {volume} {112}},\ \bibinfo
  {pages} {044140} (\bibinfo {year} {2025})}\BibitemShut {NoStop}%
\bibitem [{\citenamefont {Mattis}(1976)}]{Mattis1976}%
  \BibitemOpen
  \bibfield  {author} {\bibinfo {author} {\bibfnamefont {D.~C.}\ \bibnamefont
  {Mattis}},\ }\bibfield  {title} {\bibinfo {title} {{Solvable spin systems
  with random interactions}},\ }\href
  {https://doi.org/10.1016/0375-9601(76)90396-0} {\bibfield  {journal}
  {\bibinfo  {journal} {Phys. Lett. A}\ }\textbf {\bibinfo {volume} {56}},\
  \bibinfo {pages} {421} (\bibinfo {year} {1976})}\BibitemShut {NoStop}%
\bibitem [{\citenamefont {Oshima}\ \emph {et~al.}(2026)\citenamefont {Oshima},
  \citenamefont {Arai},\ and\ \citenamefont {Hukushima}}]{Oshima2026}%
  \BibitemOpen
  \bibfield  {author} {\bibinfo {author} {\bibfnamefont {T.}~\bibnamefont
  {Oshima}}, \bibinfo {author} {\bibfnamefont {Y.}~\bibnamefont {Arai}},\ and\
  \bibinfo {author} {\bibfnamefont {K.}~\bibnamefont {Hukushima}},\ }\bibfield
  {title} {\bibinfo {title} {{Phase transitions in a modified Ising spin glass
  model: A tensor-network-based sampling approach}},\ }\href
  {https://arxiv.org/abs/2603.18486v2} {\bibfield  {journal} {\bibinfo
  {journal} {arXiv:2603.18486v2}\ } (\bibinfo {year} {2026})}\BibitemShut
  {NoStop}%
\bibitem [{\citenamefont {Krzakala}\ and\ \citenamefont
  {Zdeborov{\'a}}(2009)}]{Krzakala2009}%
  \BibitemOpen
  \bibfield  {author} {\bibinfo {author} {\bibfnamefont {F.}~\bibnamefont
  {Krzakala}}\ and\ \bibinfo {author} {\bibfnamefont {L.}~\bibnamefont
  {Zdeborov{\'a}}},\ }\bibfield  {title} {\bibinfo {title} {{Hiding quiet
  solutions in random constraint satisfaction problems}},\ }\href
  {https://doi.org/10.1103/PhysRevLett.102.238701} {\bibfield  {journal}
  {\bibinfo  {journal} {Phys. Rev. Lett.}\ }\textbf {\bibinfo {volume} {102}},\
  \bibinfo {pages} {238701} (\bibinfo {year} {2009})}\BibitemShut {NoStop}%
\bibitem [{\citenamefont {Zdeborov{\'a}}\ and\ \citenamefont
  {Krzakala}(2016)}]{Zdeborova2016}%
  \BibitemOpen
  \bibfield  {author} {\bibinfo {author} {\bibfnamefont {L.}~\bibnamefont
  {Zdeborov{\'a}}}\ and\ \bibinfo {author} {\bibfnamefont {F.}~\bibnamefont
  {Krzakala}},\ }\bibfield  {title} {\bibinfo {title} {{Statistical physics of
  inference: thresholds and algorithms}},\ }\href
  {https://doi.org/10.1080/00018732.2016.1211393} {\bibfield  {journal}
  {\bibinfo  {journal} {Adv. Phys.}\ }\textbf {\bibinfo {volume} {65}},\
  \bibinfo {pages} {453} (\bibinfo {year} {2016})}\BibitemShut {NoStop}%
\bibitem [{\citenamefont {Zhu}\ \emph {et~al.}(2023)\citenamefont {Zhu},
  \citenamefont {Tantivasadakarn}, \citenamefont {Vishwanath}, \citenamefont
  {Trebst},\ and\ \citenamefont {Verresen}}]{Zhu2023}%
  \BibitemOpen
  \bibfield  {author} {\bibinfo {author} {\bibfnamefont {G.-Y.}\ \bibnamefont
  {Zhu}}, \bibinfo {author} {\bibfnamefont {N.}~\bibnamefont
  {Tantivasadakarn}}, \bibinfo {author} {\bibfnamefont {A.}~\bibnamefont
  {Vishwanath}}, \bibinfo {author} {\bibfnamefont {S.}~\bibnamefont {Trebst}},\
  and\ \bibinfo {author} {\bibfnamefont {R.}~\bibnamefont {Verresen}},\
  }\bibfield  {title} {\bibinfo {title} {{Nishimori's cat: Stable long-range
  entanglement from finite-depth unitaries and weak measurements}},\ }\href
  {https://doi.org/10.1103/PhysRevLett.131.200201} {\bibfield  {journal}
  {\bibinfo  {journal} {Phys. Rev. Lett.}\ }\textbf {\bibinfo {volume} {131}},\
  \bibinfo {pages} {200201} (\bibinfo {year} {2023})}\BibitemShut {NoStop}%
\bibitem [{\citenamefont {Chen}\ \emph {et~al.}(2025)\citenamefont {Chen},
  \citenamefont {Zhu}, \citenamefont {Verresen}, \citenamefont {Seif},
  \citenamefont {B{\"a}umer}, \citenamefont {Layden}, \citenamefont
  {Tantivasadakarn}, \citenamefont {Zhu}, \citenamefont {Sheldon},
  \citenamefont {Vishwanath}, \citenamefont {Trebst},\ and\ \citenamefont
  {Kandala}}]{Chen2025}%
  \BibitemOpen
  \bibfield  {author} {\bibinfo {author} {\bibfnamefont {E.~H.}\ \bibnamefont
  {Chen}}, \bibinfo {author} {\bibfnamefont {G.-Y.}\ \bibnamefont {Zhu}},
  \bibinfo {author} {\bibfnamefont {R.}~\bibnamefont {Verresen}}, \bibinfo
  {author} {\bibfnamefont {A.}~\bibnamefont {Seif}}, \bibinfo {author}
  {\bibfnamefont {E.}~\bibnamefont {B{\"a}umer}}, \bibinfo {author}
  {\bibfnamefont {D.}~\bibnamefont {Layden}}, \bibinfo {author} {\bibfnamefont
  {N.}~\bibnamefont {Tantivasadakarn}}, \bibinfo {author} {\bibfnamefont
  {G.}~\bibnamefont {Zhu}}, \bibinfo {author} {\bibfnamefont {S.}~\bibnamefont
  {Sheldon}}, \bibinfo {author} {\bibfnamefont {A.}~\bibnamefont {Vishwanath}},
  \bibinfo {author} {\bibfnamefont {S.}~\bibnamefont {Trebst}},\ and\ \bibinfo
  {author} {\bibfnamefont {A.}~\bibnamefont {Kandala}},\ }\bibfield  {title}
  {\bibinfo {title} {{Nishimori transition across the error threshold for
  constant-depth quantum circuits}},\ }\href
  {https://doi.org/10.1038/s41567-024-02696-6} {\bibfield  {journal} {\bibinfo
  {journal} {Nat. Phys.}\ }\textbf {\bibinfo {volume} {21}},\ \bibinfo {pages}
  {161} (\bibinfo {year} {2025})}\BibitemShut {NoStop}%
\bibitem [{\citenamefont {Kitatani}(1992)}]{Kitatani1992}%
  \BibitemOpen
  \bibfield  {author} {\bibinfo {author} {\bibfnamefont {H.}~\bibnamefont
  {Kitatani}},\ }\bibfield  {title} {\bibinfo {title} {The verticality of the
  ferromagnetic-spin glass phase boundary of the {$\pm J$ Ising} model in the
  {$p$-$T$} plane},\ }\href {https://doi.org/10.1143/JPSJ.61.4049} {\bibfield
  {journal} {\bibinfo  {journal} {J. Phys. Soc. Jpn.}\ }\textbf {\bibinfo
  {volume} {61}},\ \bibinfo {pages} {4049} (\bibinfo {year}
  {1992})}\BibitemShut {NoStop}%
\bibitem [{\citenamefont {Parisi}(1979)}]{Parisi1979}%
  \BibitemOpen
  \bibfield  {author} {\bibinfo {author} {\bibfnamefont {G.}~\bibnamefont
  {Parisi}},\ }\bibfield  {title} {\bibinfo {title} {{Infinite number of order
  parameters for spin-glasses}},\ }\href
  {https://doi.org/10.1103/PhysRevLett.43.1754} {\bibfield  {journal} {\bibinfo
   {journal} {Phys. Rev. Lett.}\ }\textbf {\bibinfo {volume} {43}},\ \bibinfo
  {pages} {1754} (\bibinfo {year} {1979})}\BibitemShut {NoStop}%
\bibitem [{\citenamefont {M{\'e}zard}\ \emph {et~al.}(1987)\citenamefont
  {M{\'e}zard}, \citenamefont {Parisi},\ and\ \citenamefont
  {Virasoro}}]{Mezard1987}%
  \BibitemOpen
  \bibfield  {author} {\bibinfo {author} {\bibfnamefont {M.}~\bibnamefont
  {M{\'e}zard}}, \bibinfo {author} {\bibfnamefont {G.}~\bibnamefont {Parisi}},\
  and\ \bibinfo {author} {\bibfnamefont {M.~A.}\ \bibnamefont {Virasoro}},\
  }\href@noop {} {\emph {\bibinfo {title} {{Spin Glass Theory and Beyond}}}}\
  (\bibinfo  {publisher} {World Scientific},\ \bibinfo {address} {Singapore},\
  \bibinfo {year} {1987})\BibitemShut {NoStop}%
\bibitem [{\citenamefont {Bray}\ and\ \citenamefont {Moore}(1987)}]{Bray1987}%
  \BibitemOpen
  \bibfield  {author} {\bibinfo {author} {\bibfnamefont {A.~J.}\ \bibnamefont
  {Bray}}\ and\ \bibinfo {author} {\bibfnamefont {M.~A.}\ \bibnamefont
  {Moore}},\ }\bibfield  {title} {\bibinfo {title} {Chaotic nature of the
  spin-glass phase},\ }\href {https://doi.org/10.1103/PhysRevLett.58.57}
  {\bibfield  {journal} {\bibinfo  {journal} {Phys. Rev. Lett.}\ }\textbf
  {\bibinfo {volume} {58}},\ \bibinfo {pages} {57} (\bibinfo {year}
  {1987})}\BibitemShut {NoStop}%
\bibitem [{\citenamefont {Kondor}(1989)}]{Kondor1989}%
  \BibitemOpen
  \bibfield  {author} {\bibinfo {author} {\bibfnamefont {I.}~\bibnamefont
  {Kondor}},\ }\bibfield  {title} {\bibinfo {title} {On chaos in spin
  glasses},\ }\href {https://doi.org/10.1088/0305-4470/22/5/005} {\bibfield
  {journal} {\bibinfo  {journal} {J. Phys. A}\ }\textbf {\bibinfo {volume}
  {22}},\ \bibinfo {pages} {L163} (\bibinfo {year} {1989})}\BibitemShut
  {NoStop}%
\bibitem [{\citenamefont {Ritort}(1994)}]{Ritort1994}%
  \BibitemOpen
  \bibfield  {author} {\bibinfo {author} {\bibfnamefont {F.}~\bibnamefont
  {Ritort}},\ }\bibfield  {title} {\bibinfo {title} {{Static chaos and scaling
  behavior in the spin-glass phase}},\ }\href
  {https://doi.org/10.1103/PhysRevB.50.6844} {\bibfield  {journal} {\bibinfo
  {journal} {Phys. Rev. B}\ }\textbf {\bibinfo {volume} {50}},\ \bibinfo
  {pages} {6844} (\bibinfo {year} {1994})}\BibitemShut {NoStop}%
\bibitem [{\citenamefont {Billoire}\ and\ \citenamefont
  {Coluzzi}(2003)}]{Billoire2003}%
  \BibitemOpen
  \bibfield  {author} {\bibinfo {author} {\bibfnamefont {A.}~\bibnamefont
  {Billoire}}\ and\ \bibinfo {author} {\bibfnamefont {B.}~\bibnamefont
  {Coluzzi}},\ }\bibfield  {title} {\bibinfo {title} {{Magnetic field chaos in
  the Sherrington--Kirkpatrick model}},\ }\href
  {https://doi.org/10.1103/PhysRevE.67.036108} {\bibfield  {journal} {\bibinfo
  {journal} {Phys. Rev. E}\ }\textbf {\bibinfo {volume} {67}},\ \bibinfo
  {pages} {036108} (\bibinfo {year} {2003})}\BibitemShut {NoStop}%
\bibitem [{\citenamefont {Aguilar-Janita}\ \emph {et~al.}(2024)\citenamefont
  {Aguilar-Janita}, \citenamefont {Franz}, \citenamefont {Martin-Mayor},
  \citenamefont {Moreno-Gordo}, \citenamefont {Parisi}, \citenamefont
  {Ricci-Tersenghi},\ and\ \citenamefont {Ruiz-Lorenzo}}]{AguilarJanita2024}%
  \BibitemOpen
  \bibfield  {author} {\bibinfo {author} {\bibfnamefont {M.}~\bibnamefont
  {Aguilar-Janita}}, \bibinfo {author} {\bibfnamefont {S.}~\bibnamefont
  {Franz}}, \bibinfo {author} {\bibfnamefont {V.}~\bibnamefont {Martin-Mayor}},
  \bibinfo {author} {\bibfnamefont {J.}~\bibnamefont {Moreno-Gordo}}, \bibinfo
  {author} {\bibfnamefont {G.}~\bibnamefont {Parisi}}, \bibinfo {author}
  {\bibfnamefont {F.}~\bibnamefont {Ricci-Tersenghi}},\ and\ \bibinfo {author}
  {\bibfnamefont {J.~J.}\ \bibnamefont {Ruiz-Lorenzo}},\ }\bibfield  {title}
  {\bibinfo {title} {{Small field chaos in spin glasses: Universal predictions
  from the ultrametric tree and comparison with numerical simulations}},\
  }\href {https://doi.org/10.1073/pnas.2404973121} {\bibfield  {journal}
  {\bibinfo  {journal} {Proc. Natl. Acad. Sci. USA}\ }\textbf {\bibinfo
  {volume} {121}},\ \bibinfo {pages} {e2404973121} (\bibinfo {year}
  {2024})}\BibitemShut {NoStop}%
\bibitem [{\citenamefont {Fisher}\ and\ \citenamefont
  {Huse}(1986)}]{Fisher1986}%
  \BibitemOpen
  \bibfield  {author} {\bibinfo {author} {\bibfnamefont {D.~S.}\ \bibnamefont
  {Fisher}}\ and\ \bibinfo {author} {\bibfnamefont {D.~A.}\ \bibnamefont
  {Huse}},\ }\bibfield  {title} {\bibinfo {title} {Ordered phase of short-range
  {Ising} spin-glasses},\ }\href {https://doi.org/10.1103/PhysRevLett.56.1601}
  {\bibfield  {journal} {\bibinfo  {journal} {Phys. Rev. Lett.}\ }\textbf
  {\bibinfo {volume} {56}},\ \bibinfo {pages} {1601} (\bibinfo {year}
  {1986})}\BibitemShut {NoStop}%
\bibitem [{\citenamefont {Fisher}\ and\ \citenamefont
  {Huse}(1988)}]{Fisher1988}%
  \BibitemOpen
  \bibfield  {author} {\bibinfo {author} {\bibfnamefont {D.~S.}\ \bibnamefont
  {Fisher}}\ and\ \bibinfo {author} {\bibfnamefont {D.~A.}\ \bibnamefont
  {Huse}},\ }\bibfield  {title} {\bibinfo {title} {{Equilibrium behavior of the
  spin-glass ordered phase}},\ }\href {https://doi.org/10.1103/PhysRevB.38.386}
  {\bibfield  {journal} {\bibinfo  {journal} {Phys. Rev. B}\ }\textbf {\bibinfo
  {volume} {38}},\ \bibinfo {pages} {386} (\bibinfo {year} {1988})}\BibitemShut
  {NoStop}%
\bibitem [{\citenamefont {Imry}\ and\ \citenamefont {Ma}(1975)}]{Imry1975}%
  \BibitemOpen
  \bibfield  {author} {\bibinfo {author} {\bibfnamefont {Y.}~\bibnamefont
  {Imry}}\ and\ \bibinfo {author} {\bibfnamefont {S.-K.}\ \bibnamefont {Ma}},\
  }\bibfield  {title} {\bibinfo {title} {{Random-field instability of the
  ordered state of continuous symmetry}},\ }\href
  {https://doi.org/10.1103/PhysRevLett.35.1399} {\bibfield  {journal} {\bibinfo
   {journal} {Phys. Rev. Lett.}\ }\textbf {\bibinfo {volume} {35}},\ \bibinfo
  {pages} {1399} (\bibinfo {year} {1975})}\BibitemShut {NoStop}%
\bibitem [{\citenamefont {Imbrie}(1985)}]{Imbrie1985}%
  \BibitemOpen
  \bibfield  {author} {\bibinfo {author} {\bibfnamefont {J.~Z.}\ \bibnamefont
  {Imbrie}},\ }\bibfield  {title} {\bibinfo {title} {{The ground state of the
  three-dimensional random-field Ising model}},\ }\href
  {https://doi.org/10.1007/BF01220505} {\bibfield  {journal} {\bibinfo
  {journal} {Commun. Math. Phys.}\ }\textbf {\bibinfo {volume} {98}},\ \bibinfo
  {pages} {145} (\bibinfo {year} {1985})}\BibitemShut {NoStop}%
\bibitem [{\citenamefont {Bricmont}\ and\ \citenamefont
  {Kupiainen}(1988)}]{BricmontKupiainen1988}%
  \BibitemOpen
  \bibfield  {author} {\bibinfo {author} {\bibfnamefont {J.}~\bibnamefont
  {Bricmont}}\ and\ \bibinfo {author} {\bibfnamefont {A.}~\bibnamefont
  {Kupiainen}},\ }\bibfield  {title} {\bibinfo {title} {{Phase transition in
  the 3d random field Ising model}},\ }\href
  {https://doi.org/10.1007/BF01224901} {\bibfield  {journal} {\bibinfo
  {journal} {Commun. Math. Phys.}\ }\textbf {\bibinfo {volume} {116}},\
  \bibinfo {pages} {539} (\bibinfo {year} {1988})}\BibitemShut {NoStop}%
\bibitem [{\citenamefont {Aizenman}\ and\ \citenamefont
  {Wehr}(1989)}]{AizenmanWehr1989}%
  \BibitemOpen
  \bibfield  {author} {\bibinfo {author} {\bibfnamefont {M.}~\bibnamefont
  {Aizenman}}\ and\ \bibinfo {author} {\bibfnamefont {J.}~\bibnamefont
  {Wehr}},\ }\bibfield  {title} {\bibinfo {title} {{Rounding of first-order
  phase transitions in systems with quenched disorder}},\ }\href
  {https://doi.org/10.1103/PhysRevLett.62.2503} {\bibfield  {journal} {\bibinfo
   {journal} {Phys. Rev. Lett.}\ }\textbf {\bibinfo {volume} {62}},\ \bibinfo
  {pages} {2503} (\bibinfo {year} {1989})}\BibitemShut {NoStop}%
\bibitem [{\citenamefont {Aizenman}\ and\ \citenamefont
  {Wehr}(1990)}]{AizenmanWehr1990}%
  \BibitemOpen
  \bibfield  {author} {\bibinfo {author} {\bibfnamefont {M.}~\bibnamefont
  {Aizenman}}\ and\ \bibinfo {author} {\bibfnamefont {J.}~\bibnamefont
  {Wehr}},\ }\bibfield  {title} {\bibinfo {title} {{Rounding effects of
  quenched randomness on first-order phase transitions}},\ }\href
  {https://doi.org/10.1007/BF02096933} {\bibfield  {journal} {\bibinfo
  {journal} {Commun. Math. Phys.}\ }\textbf {\bibinfo {volume} {130}},\
  \bibinfo {pages} {489} (\bibinfo {year} {1990})}\BibitemShut {NoStop}%
\bibitem [{\citenamefont {Hartmann}\ and\ \citenamefont
  {Young}(2001)}]{Hartmann2001}%
  \BibitemOpen
  \bibfield  {author} {\bibinfo {author} {\bibfnamefont {A.~K.}\ \bibnamefont
  {Hartmann}}\ and\ \bibinfo {author} {\bibfnamefont {A.~P.}\ \bibnamefont
  {Young}},\ }\bibfield  {title} {\bibinfo {title} {{Lower critical dimension
  of Ising spin glasses}},\ }\href {https://doi.org/10.1103/PhysRevB.64.180404}
  {\bibfield  {journal} {\bibinfo  {journal} {Phys. Rev. B}\ }\textbf {\bibinfo
  {volume} {64}},\ \bibinfo {pages} {180404(R)} (\bibinfo {year}
  {2001})}\BibitemShut {NoStop}%
\bibitem [{\citenamefont {J\"org}\ \emph {et~al.}(2006)\citenamefont {J\"org},
  \citenamefont {Lukic}, \citenamefont {Marinari},\ and\ \citenamefont
  {Martin}}]{Jorg2006}%
  \BibitemOpen
  \bibfield  {author} {\bibinfo {author} {\bibfnamefont {T.}~\bibnamefont
  {J\"org}}, \bibinfo {author} {\bibfnamefont {J.}~\bibnamefont {Lukic}},
  \bibinfo {author} {\bibfnamefont {E.}~\bibnamefont {Marinari}},\ and\
  \bibinfo {author} {\bibfnamefont {O.~C.}\ \bibnamefont {Martin}},\ }\bibfield
   {title} {\bibinfo {title} {Strong universality and algebraic scaling in
  two-dimensional {Ising} spin glasses},\ }\href
  {https://doi.org/10.1103/PhysRevLett.96.237205} {\bibfield  {journal}
  {\bibinfo  {journal} {Phys. Rev. Lett.}\ }\textbf {\bibinfo {volume} {96}},\
  \bibinfo {pages} {237205} (\bibinfo {year} {2006})}\BibitemShut {NoStop}%
\bibitem [{\citenamefont {Okuyama}\ \emph {et~al.}(2026)\citenamefont
  {Okuyama}, \citenamefont {Ohzeki},\ and\ \citenamefont
  {Nishimori}}]{Okuyama2026}%
  \BibitemOpen
  \bibfield  {author} {\bibinfo {author} {\bibfnamefont {M.}~\bibnamefont
  {Okuyama}}, \bibinfo {author} {\bibfnamefont {M.}~\bibnamefont {Ohzeki}},\
  and\ \bibinfo {author} {\bibfnamefont {H.}~\bibnamefont {Nishimori}},\
  }\bibfield  {title} {\bibinfo {title} {{Spin-glass order in a field requires
  magnetic-field chaos in XY gauge glasses for $d\leq 4$}},\ }\href
  {https://arxiv.org/abs/2611.14109} {\bibfield  {journal} {\bibinfo  {journal}
  {arXiv:2611.14109}\ } (\bibinfo {year} {2026})}\BibitemShut {NoStop}%
\bibitem [{\citenamefont {Mermin}\ and\ \citenamefont
  {Wagner}(1966)}]{MerminWagner1966}%
  \BibitemOpen
  \bibfield  {author} {\bibinfo {author} {\bibfnamefont {N.~D.}\ \bibnamefont
  {Mermin}}\ and\ \bibinfo {author} {\bibfnamefont {H.}~\bibnamefont
  {Wagner}},\ }\bibfield  {title} {\bibinfo {title} {{Absence of ferromagnetism
  or antiferromagnetism in one- or two-dimensional isotropic Heisenberg
  models}},\ }\href {https://doi.org/10.1103/PhysRevLett.17.1133} {\bibfield
  {journal} {\bibinfo  {journal} {Phys. Rev. Lett.}\ }\textbf {\bibinfo
  {volume} {17}},\ \bibinfo {pages} {1133} (\bibinfo {year}
  {1966})}\BibitemShut {NoStop}%
\bibitem [{\citenamefont {Dario}\ \emph {et~al.}(2024)\citenamefont {Dario},
  \citenamefont {Harel},\ and\ \citenamefont {Peled}}]{Dario2024}%
  \BibitemOpen
  \bibfield  {author} {\bibinfo {author} {\bibfnamefont {P.}~\bibnamefont
  {Dario}}, \bibinfo {author} {\bibfnamefont {M.}~\bibnamefont {Harel}},\ and\
  \bibinfo {author} {\bibfnamefont {R.}~\bibnamefont {Peled}},\ }\bibfield
  {title} {\bibinfo {title} {{Quantitative disorder effects in low-dimensional
  spin systems}},\ }\href {https://doi.org/10.1007/s00220-024-05081-9}
  {\bibfield  {journal} {\bibinfo  {journal} {Commun. Math. Phys.}\ }\textbf
  {\bibinfo {volume} {405}},\ \bibinfo {pages} {212} (\bibinfo {year}
  {2024})}\BibitemShut {NoStop}%
\bibitem [{\citenamefont {Nishimori}(2026)}]{Nishimori2026pspin}%
  \BibitemOpen
  \bibfield  {author} {\bibinfo {author} {\bibfnamefont {H.}~\bibnamefont
  {Nishimori}},\ }\bibfield  {title} {\bibinfo {title} {{Reentrance and
  temperature chaos in the $p$-spin Ising spin glass}},\ }\href
  {https://doi.org/10.1103/1xrx-rsxn} {\bibfield  {journal} {\bibinfo
  {journal} {Phys. Rev. E}\ }\textbf {\bibinfo {volume} {114}},\ \bibinfo
  {pages} {044127} (\bibinfo {year} {2026})}\BibitemShut {NoStop}%
\end{thebibliography}%

\end{document}